\documentclass[epj]{svjour}
\usepackage{graphicx}
\usepackage{bm}
\usepackage{float}
\usepackage{amssymb}
\newcommand{\amev}{\textit{A} MeV}
\begin{document}
\title{Experimental effects on dynamics and thermodynamics in \\
nuclear reactions on the symmetry energy as seen by the CHIMERA 4$\bm{\pi}$
detector}
\author{E. De Filippo\inst{1} \and
        A. Pagano\inst{1}
}
\institute{INFN, sez. di Catania, via S. Sofia 64, 95123 Catania, Italy}
\date{Received: \today } 
\abstract {
Heavy ion collisions have been widely used in the last decade to constraint the
parameterizations of the symmetry energy term of nuclear equation of state (EOS) for
asymmetric nuclear matter as a function of baryonic density. In the Fermi energy domain
one is faced with variations of the density within a narrow range of values
around the saturation density $\rho_0$=0.16 fm$^{-3}$ down towards sub-saturation densities.
The experimental observables which are sensitive to the symmetry energy are constructed
starting from the detected light particles, clusters and heavy fragments that, in heavy ion collisions,
are generally produced by different emission mechanisms at different stages and time scales of the
reaction. In this review the effects of dynamics and thermodynamics on the symmetry energy
in nuclear reactions are discussed and characterized using an overview of the data taken so far with
the CHIMERA multi-detector array.
}
\titlerunning{Experimental effects on dynamics...}
\maketitle
\section{Introduction}
\label{intro}
Energetic nucleus-nucleus collisions are very efficient tools to produce heavy fragments, clusters of
intermediate mass (IMF) and light particles. Large dissipations of both kinetic energy and angular
momentum are also involved in the reaction. Experimental signatures indicate that excited nuclear
systems are produced by different competing mechanisms. Dynamical emission, pre-equilibrium and
statistical decay of excited systems at equilibrium are at the origin of many phenomena:
break-up and projectile or target fragmentation, neck emission, fission, multifragmentation.
At the \emph{Fermi beam energies} (between 10 and 100 \emph{A} MeV) all these processes can
coexist in a complex way due to the strong competition between mean field dynamics and two body in
medium interactions, as a function of the incident beam energy, impact parameter and the time scale
of the reaction dynamics.

Disentangling among the different  mechanisms in the particle emission still it remains
a real challenge for experimental physics. Particle emissions from the early phases of the dynamical
evolution (few fm/c) up to later stages (up to several hundreds of fm/c) of statistical
decays from excited systems have been  measured.
In these studies it was necessary to unfold the various processes according to their characteristic
time scale.
This aspect is particularly important when the isospin degree of freedom is investigated
in order to probe the behavior of the symmetry energy of the nuclear equation of state (EOS) in the
density range where reaction simulations indicate that particles and fragments have been originated.

Generally, to reach this goal, projectiles and targets with large isospin asymmetries
$I = (N-Z)/(N+Z)$ are used, where $N$ and $Z$ are respectively the neutron and proton numbers, using stable neutron-rich, neutron-poor or radioactive beams (for a review see \cite{bao08,bao01}).
We notice that the density dependence of symmetry energy is not only a key ingredient for dynamical models of heavy ion collisions but also it is important in nuclear structure studies of exotic nuclei and for realistic astrophysical predictions on the structure of neutron stars \cite{tsa12,lat07}.

Different experimental observables have been used to constraint the density dependence of the symmetry energy at sub-saturation densities: isospin diffusion and equilibration \cite{tsa09,gal09,gal09_1}, neutron to proton ratio \cite{fam06}, transverse collective flow of light charged particles \cite{koh11}, ratio of fragments yields and isoscaling \cite{xu00,tsaPRL01}, heavy residue production in semi-central collisions \cite{amo09}, isospin migration of neutrons and protons between the low density ``neck'' region and
denser matter in the proximity \cite{def12}, cluster formation at very low densities
\cite{nat10,wad12}. Such studies have to be further extended by investigations at supra-saturation
density in order to obtain non contradicting results when the full parametrization of EOS is applied to
describe in a coherent way both astrophysical observations and heavy ion phenomenology \cite{tra12}.
Recently, the CHIMERA detector was
involved in a specific experiment at GSI \cite{rus13} aimed to study elliptic flow observable for
neutron and protons taken in coincidence with charged particles \cite{rus11}.

Typical observables of isospin physics are constructed starting from the neutron to proton ratio
$N/Z$ value of the detected light particles and fragments and thus they reflect the dynamical evolution
of the physical processes. Anyway the primary information carried by the isospin contents of the
particles can be distorted at different time scales along the pathway of the reaction: from the early
dynamical phase toward the pre-equilibrium emission that happens in a time scale before the full
equilibrium of the isospin degree of freedom will be achieved, until to
the sequential decay mechanism from highly excited nuclei, unavoidably present in the last
stage of the reaction.
From the experimental point of view, the use of 4$\pi$ detectors, occasionally coupled to
specialized devices like particle-particle correlators or segmented neutron detectors
\cite{fam06,mar13} is a powerful method to collect events with high particle multiplicity
at the Fermi energies.

\begin{figure}[b]
\includegraphics[width=0.50\textwidth]{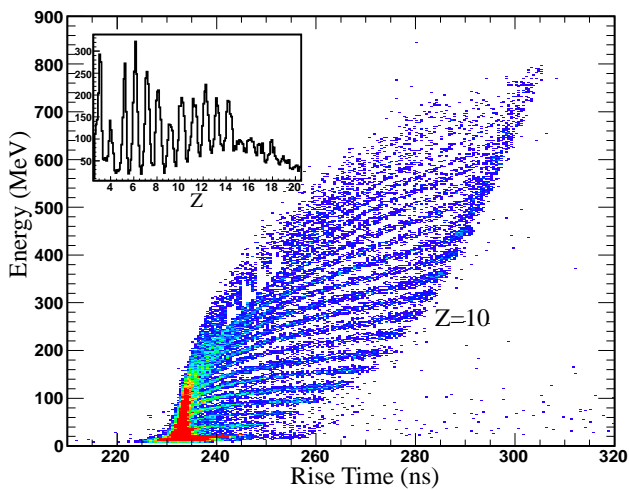}
\caption{Rise-time  vs. kinetic energy
for particles stopping in a silicon detector at around 13.5$^o$ for the reaction $^{64}$Ni+
$^{124}$Sn at 35 \amev. The inset shows the quality of charge identification obtained
in the experiment.}
\label{fig1}
\end{figure}

In this review we focus our attention upon effects on dynamics and thermodynamics in
nuclear reactions on the symmetry energy as seen by the CHIMERA detector with particular emphasis
on time scale of the reaction dynamics. This powerful device combines good time-of-flight measurements in a 4$\pi$ configuration with low threshold and isotopic identification of light fragments. The reaction dynamics and the isospin degree of freedom have been studied in a broad range of projectile-target combination at LNS superconducting cyclotron in Catania, looking at neck fragmentation and isospin diffusion, nuclear multifragmentation in central collisions, compound nucleus formation and decay, odd-even effects (\emph{staggering}) in fragment production, incomplete fusion reactions,
statistical and dynamical fission, exotic break-up of heavy systems.

\begin{figure}[t]
\includegraphics[width=0.50\textwidth]{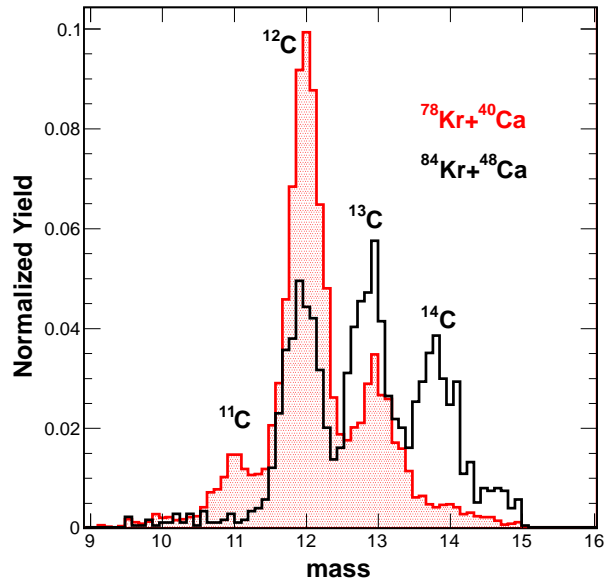}
\caption{Mass distributions (normalized to their respective area) for charge Z=6, obtained integrating the yield of all detectors at 15${^o}$ (Ring 12) for the two reactions $^{78}Kr+^{40}Ca$ (red hashed histogram)
and $^{84}Kr+^{48}Ca$ (black histogram) at 10 \amev. }
\label{figisotop}
\end{figure}

\section{The CHIMERA multidetector in brief}
\label{sec:1}
The CHIMERA multidetector (see ref. \cite{pag04,pag12} for a review), in its standard configuration, is constituted by 1192 telescopes arranged in 35 rings in a full 2$\pi$ azimuthal symmetry around the beam axis, covering the polar angle between 1$^o$ and 176$^o$. A single detection cell is constituted of a planar n-type silicon detector ($\sim300\mu m$ thickness) followed by a CsI(Tl) scintillator of thickness variable from 12 cm at forward angles to 3 cm at backward angles. The cells placed between 30$^o$ and
176$^o$ are arranged in stainless steel spherical structure (40 cm radius).

\begin{figure}[t]
\includegraphics[width=0.50\textwidth]{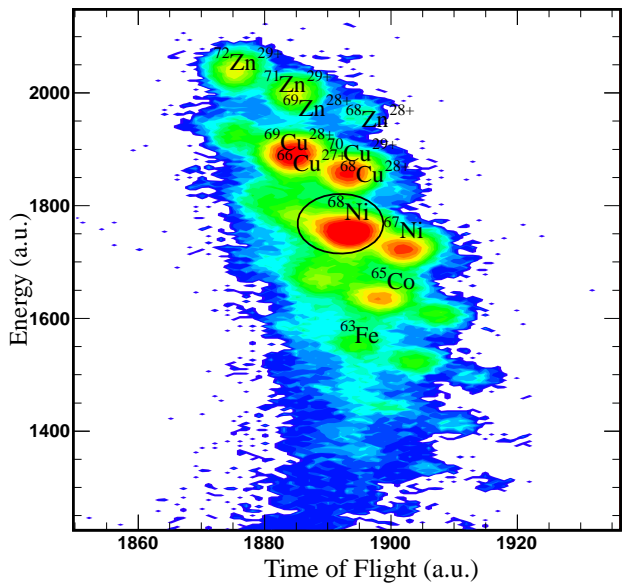}
\caption{Bidimensional plot of Time-of-Flight vs. loss of energy in a strip of a DSSSD detector
of the CHIMERA tagging system during runs for production of $^{68}Ni$ beams (indicated by a circle)
at INFN-LNS.
}
\label{figfribs}
\end{figure}

\begin{figure}[bh]
\includegraphics[width=0.50\textwidth]{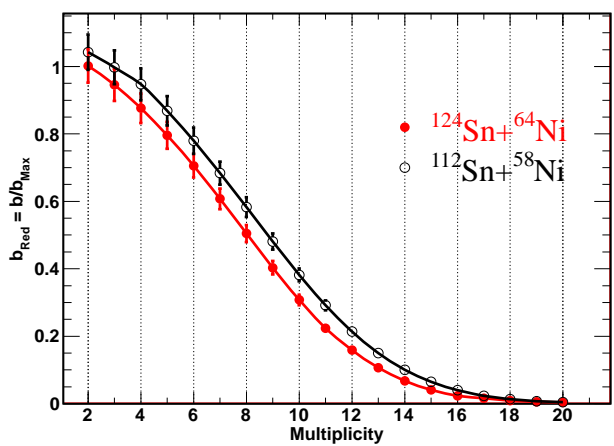}
\caption{Reduced impact parameter $b_{red}$ plotted as function of the total charged particles multiplicity for the $^{124}$Sn+$^{64}$Ni  and $^{112}$Sn+$^{58}$Ni at 35 \amev\ reactions.}
\label{fig4}
\end{figure}

Four identification techniques are currently used: i) fragments punching through the silicon detectors are identified in charge $Z$ by the ${\Delta}E-E$ technique and light fragments (up to Z$\le$9) are isotopically identified \cite{len02}. ii) Light charged particles (Z$\le$3) are isotopically identified in the CsI(Tl) by pulse-shape discrimination \cite{ald02}. iii) Velocity determination and mass evaluation (for particles stopping in the silicon detectors) are performed via the time-of-flight measurement using the timing signal delivered by the silicon detectors (30\% constant fraction discrimination is employed) with respect to the high-frequency signal (RF) of the cyclotron pulsed beam.
Average timing resolution of 1 ns has been obtained (when the timing resolution of the pulsed beam had values of 0.8 ns or better) iv) Finally, fragments stopped in the silicon detectors are identified
in charge by applying the pulse-shape method (PSD) through rise-time measurement of the
signals \cite{ald05}. Fig. \ref{fig1} shows an example of the charge identification obtained in the reaction $^{64}$Ni+$^{124}$Sn at 35 \amev\ with the PSD method for fragments which were stopped in silicon detectors. A good identification of the atomic number $Z$ up to Z=20 is obtained. Fig. \ref{figisotop} shows an example of isotopic identification for particles punching-through the silicon detector. The mass distribution for Carbon isotopes is obtained for all detectors at polar angle of 15$^o$ for the two reactions
$^{78}Kr$ + $^{40}Ca$ (neutron poor) and $^{86}Kr$ + $^{48}Ca$ (neutron rich) studied at 10 \amev\ \cite{pir13}. A good isotopic identification of light particles is a basic experimental
condition for the isospin physics investigations.
The availability at INFN-LNS of in-flight fragmentation beams \cite{rac08} has opened new
interesting opportunities, in particular with the possibility to obtain, in the intermediate energies regime, neutron-rich or neutron-poor radioactive beams. In Fig. \ref{figfribs}
is shown the recent production of a $^{68}$Ni radioactive beam by projectile fragmentation
of a primary $^{70}$Zn (40 \amev) beam.
The ion beam is identified event-by-event in charge, mass and energy by using the CHIMERA tagging system consisting of two double side silicon strip detectors (DSSSD) and a large surface microchannel plate (MCP) detector \cite{aco13}.

\begin{figure*}[th]
\begin{center}
\includegraphics[width=0.70\textwidth]{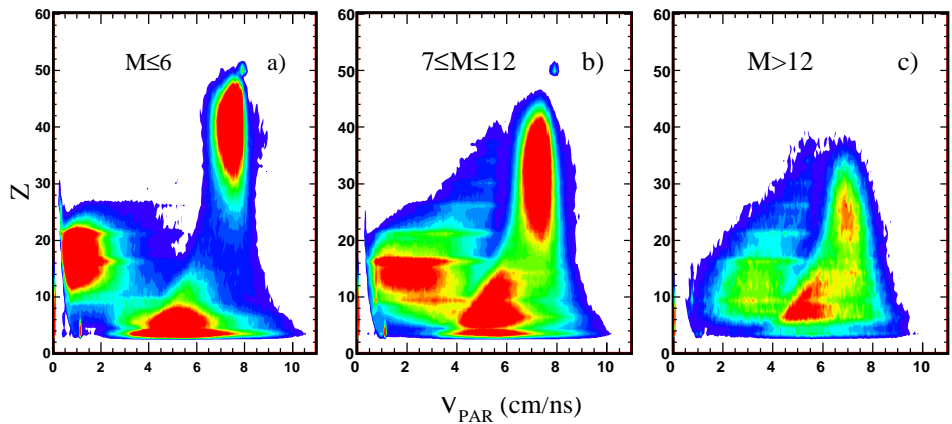}
\end{center}
\caption{Atomic number $Z$ of the three heaviest fragments in the event
for the $^{124}$Sn+$^{64}$Ni at 35 \amev\ reaction is plotted as a function of V$_{PAR}$,
for three different bins of the total charged particles multiplicity $M$.}
\label{fig5}
\end{figure*}

\subsection{An introductory view of event selection and characterization}
\label{sec:11}
The off-line analysis performed with a 4$\pi$ detector requires proper methods or criteria for selecting and sampling the collected events, before any physical analysis will be started. Among different possible selection criteria we underline the commonly accepted ones: 1) by rejecting bad detected events, including empirical procedures to remove from the analysis those detectors showing very poor response functions; 2) by identifying ``complete event'', rejecting those events where only a small percentage of the total charge and total linear momentum associated with the projectile-target entrance channel systems are collected;
and 3) constraining events as a function of useful global observables
correlated with the impact parameter of the reaction, in a event-by-event analysis.
This last aspect is particularly important because the reaction mechanism pattern strongly
depends upon the impact parameter selection.
Generally, the multiplicity of the detected charged particles or their transverse energy are used as
global observables when interested to select from peripheral to semi-central collisions on large
impact parameter ranges; the Cavata method \cite{cav90} can be used to estimate the impact
parameter from the selected observable.

Fig. \ref{fig4} shows the correlation between the total charged particles multiplicity and the reduced impact parameter $b_{red}=b/b_{Max}$ for the two reactions $^{124}Sn + ^{64}Ni$ and $^{112}Sn + ^{58}Ni$
at 35 \amev\ \cite{rusdoc} where b$_{Max}$ corresponds to the total geometrical cross section. Note that at a fixed $b_{red}$ the neutron poor system produces a higher (one unit) multiplicity. In fact, as confirmed by calculations \cite{bar05,pla08}, in the neutron poor system the dynamical and pre-equili\-brium phases lead to more light charged particles (mainly protons) emission than in neutron rich one where, in contrast, neutron emission is favored.

An example of a typical event selection in CHIMERA using the charged particles multiplicity is shown
in Fig. \ref{fig5} for the inverse kinematics reaction $^{124}$Sn+$^{64}$Ni at 35 \amev\  \cite{def12,pla08,def05}, selecting the events for which the total charge Z$_{tot}$ and
the total parallel momentum of the detected particles are larger than 70\% of the total charge
and projectile momentum of the system. The atomic number $Z$ of the
three heaviest fragments in the events versus the parallel velocity $V_{PAR}$ are plotted
for three bins of the total charged particle multiplicity $M$, going from semi-peripheral reactions
($M\le 6$) to the most dissipative collisions ($M\geq 12$).

Fig. \ref{fig5}a) shows that three groups of fragments can be separated:
fragments originated from a projectile (PLF, $V_{PAR}\approx V_{proj}$ = 8 cm/ns)
or target (TLF, $V_{PAR} \approx$ 1 cm/ns) remnants, and the class of intermediate velocity
fragments at mid-rapidity. By increasing the degree of dissipation of
the reaction, two sources of emission remain evident, indicating a clear
persistency of the binary (PLF,TLF) character of the reaction.
\begin{figure}[th]
\includegraphics[width=0.51\textwidth]{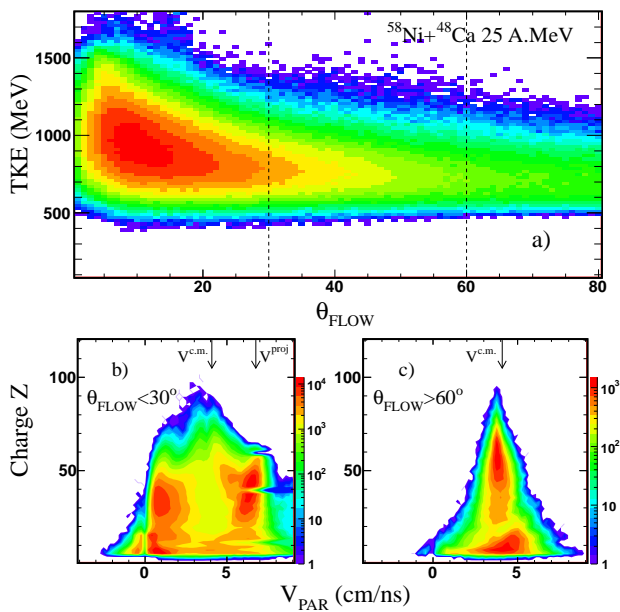}
\caption{a) $\theta_{FLOW}$ angle as a function of the
total kinetic energy of the emitted particles for the reaction $^{58}Ni$+$^{48}Ca$
at 25 \amev; b) the atomic number $Z$ of the emitted fragments
in the event is plotted as a function of V$_{PAR}$ for $\theta_{FLOW}<30^o$; c) the same for
$\theta_{FLOW} > 60^o$. }
\label{fig6}
\end{figure}

More complex methods, as principal component analysis PCA) \cite{ger04} or sorting methods based on observables related to the momentum shape of the events are needed when small cross-sections
in central collisions have to be selected (for example for characterization of
multifragmentation source  from a single fused source) \cite{fra01,bor02}. Fig. \ref{fig6}a) shows for the reaction $^{58}Ni + ^{48}Ca$ at 25 \amev, studied with the CHIMERA detector \cite{fra12}, the total kinetic energy (TKE) of detected products in almost complete events as a function of the $\theta_{FLOW}$ angle, in a ``Wilczy\'nski-like'' diagram. This last is constructed event-by-event, as the angle between the beam direction and the main emission direction of particles determined from the cartesian coordinates of the measured linear momenta \cite{ger04,fra01}. For peripheral and semi-peripheral collisions, where the events keep memory of the binary character of the reaction, the shape is elliptic and with
small values of $\theta_{FLOW}$, while for more central collisions a spherical shape is predicted corresponding to larger values (up to 90 degrees) of the  $\theta_{FLOW}$ angle. Figs. \ref{fig6}b) and \ref{fig6}c) show that at increasing values of $\theta_{FLOW}$, central collisions
characterized by a single emission source centered at
V$_{c.m.}$ are dominating with respect to the binary dissipative collisions.

\section{``Neck'' dynamics}
\label{sec:3}
At medium energy, semi-peripheral events in heavy ion collisions are characterized
by binary reactions where projectile and target nuclei experience a substantial overlap of matter (participant region). In these reactions, beside an abundant production of excited PLF and TLF nuclei,
a complex dynamical rearrangements of nucleons
in the participant region within a short time scale ($<$50 fm/c) takes place \cite{mon94}.
Transport model calculations indicate the formation of a dynamical neck-like structure at sub-normal baryonic density \cite{bar04} that is at the origin of the
light fragment (IMF) midrapidity emission \cite{dem96}.
The peculiarities of both kinematical properties and neutron enrichment of light particles and fragments originated at midrapidity carry important information on reaction dynamics and symmetry energy \cite{koh11,def12,dito06}. Thus, IMFs are mainly the remnants of a neck-like structure formed,
on a very short time scale, during the stage of reseparation of the PLF and TLF.
In the context of transport model calculations the observed neutron enrichment (neutron asymmetry) of the IMFs is generally understood by an isospin density dependent transport migration of neutrons against protons due to the density gradient \cite{barP05} between the dilute neck region and
the PLF and TLF residues.

Adopting this interpretation, much effort is devoted nowadays to determine the density evolution of the mid-rapidity source \cite{deffi13}. On the other hand light charged particles and
IMFs emitted at mid-rapidity can be produced by different reaction mechanisms
with different chronology, like pre-equilibrium emission from deformed PLF and TLF
nuclei or sequential emission in a later stage of the reaction
\cite{koh12,gin02,pia02,hud12}, as a result of a delicate competition between one body
dissipation mechanism and two body in medium interaction.

One can ask the questions:  i) can the neck fragmentation of light IMFs and  the dynamical asymmetrical fission of heavy fragments \cite{hud12,rus10} be considered as two distinct reaction mechanisms (see Section \ref{sec:4}) ? ii) is there a continuous transition from neck fragmentation towards the high multiplicity multifragmentation observed by increasing the centrality of the collision \cite{bar12} ? iii) can the experimentally observed $N/Z$ enrichment of the neck fragments unambiguously attributed to a symmetry energy effect of the nuclear equation of state (EOS) of finite matter at subsaturation density \cite{def12,sob00,pia06}?

\begin{figure}[th]
\includegraphics[width=0.50\textwidth]{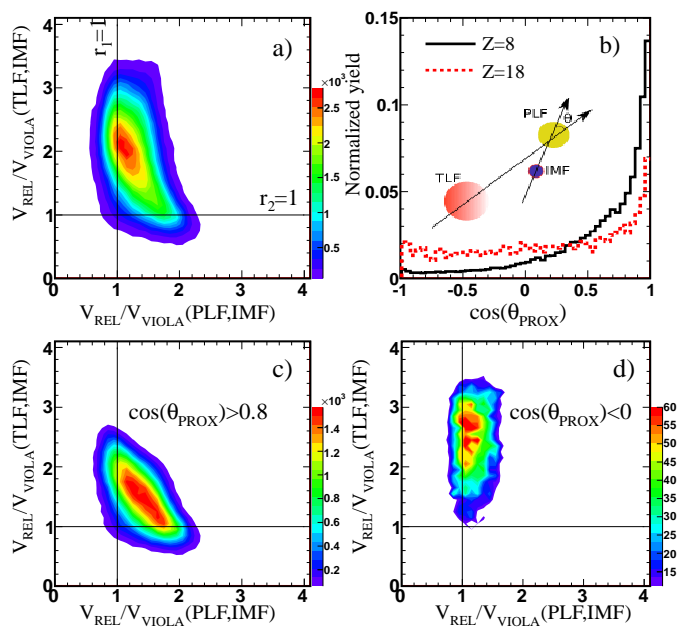}
\caption{For the $^{124}Sn+^{64}Ni$ reaction at 35 \amev: a) correlations between relative velocities $V_{rel}/V_{Viola}$ of the three biggest fragments in the event for IMFs of charge $3\leq Z \leq 18$.
b) Distribution of $\cos(\theta_{PROX})$ for Z=8 (black line) and Z=18 IMFs (red line). c) as a) with the condition $\cos(\theta_{PROX})>0.8$ and d) $\cos(\theta_{PROX})<0$. Adapted from Ref. \cite{def12}.}
\label{fig7}
\end{figure}

\subsection{Kinematic correlations and emission timescales}
\label{sec:31}
The capability of CHIMERA to detect fragments in a broad range of kinetic energies (including the slow moving target-like residues) has been applied to establish a clear new methodology to correlate the isotopic composition of IMFs emitted at midrapidity with the emission time scale in semi-peripheral collisions. It was shown that the experimental capability to select fragments formed on a broad range
of interaction timescales is extremely important for the isospin dynamics studies.
The method was applied to ternary reactions produced in semi-peripheral collisions \cite{def05,wil05}
and it is based on i) the evaluation of fragment-fragment relative velocities in a three body kinematical analysis ii) measuring the angular distributions of fragments in order to evaluate their \emph{alignment}'s properties with respect to a well defined separation axis.

\begin{figure}[th]
\includegraphics[width=0.50\textwidth]{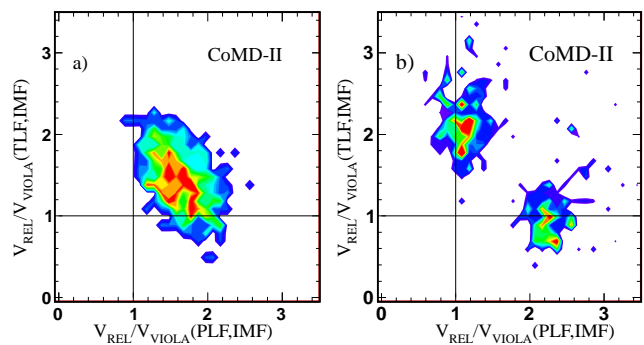}
\caption{CoMD-II calculations for the $^{124}Sn+^{64}Ni$ reaction at 35 \amev: a) IMFs emitted in the midrapidy region between 2.5 and 5 cm/ns; b) the same for IMFs velocities outside the range 2.5-5 cm/ns. The color scale is in arbitrary unit. Adapted from Ref. \cite{pap07}.}
\label{fig8qmd}
\end{figure}

Fig. \ref{fig7}a) shows an example of such a kind of analysis for the $^{124}Sn+^{64}Ni$ at 35 \amev.
For all IMFs with charges $3\leq Z \leq 18$ the relative velocities $V_{REL}$ for the couples of fragments (PLF,IMF) and (TLF,IMF) have been evaluated according to the ratios:  $r_1=V_{REL}/V_{Viola}\-(PLF,IMF)$ and $r_2=V_{REL}/V_{Viola}\-(TLF,IMF)$, where $V_{Viola}$ is the velocity due to the Coulomb repulsion for a binary splitting, as provided by the Viola systematics \cite{hin87}.
The values $r_1=1$ and $r_2=1$ indicate the loci for a sequential decay (or statistical fission) of IMFs from, respectively, the PLF and TLF sources; values of $r_1$ and $r_2$ simultaneously larger than unity indicate IMFs of dynamical origin as in a prompt ternary division. This new kind of Wilczy\'nski-like plot \cite{pag04,wil05} has been used to calibrate the timescale of IMFs emission in semi-peripheral collisions under the assumption of collinear motion of IMFs along the PLF-TLF relative
velocity direction. Firstly applied to the experimental data of the reaction $^{124}Sn$ + $^{64}Ni$ \cite{def05} simple kinematical evaluations have shown a well defined chronology: light IMFs (Z$<9$) are
emitted either on a short timescale (within 50 fm/c) with a prompt neck-rupture mechanism or
sequentially ($>$120 fm/c) after the re-separation of the binary PLF-TLF system.
Heavier fragments ($Z\ge9$) are emitted on a longer timescale and
their pattern covers a time scale ranging from a fast (on the time scale of $\approx$ 300 fm/c)
non-equilibrated fission-like splitting to a fully equilibrated fission process
of longer timescale \cite{rus10,deff05} (see Section \ref{sec:4}).

\begin{figure}[h]
\includegraphics[width=0.50\textwidth]{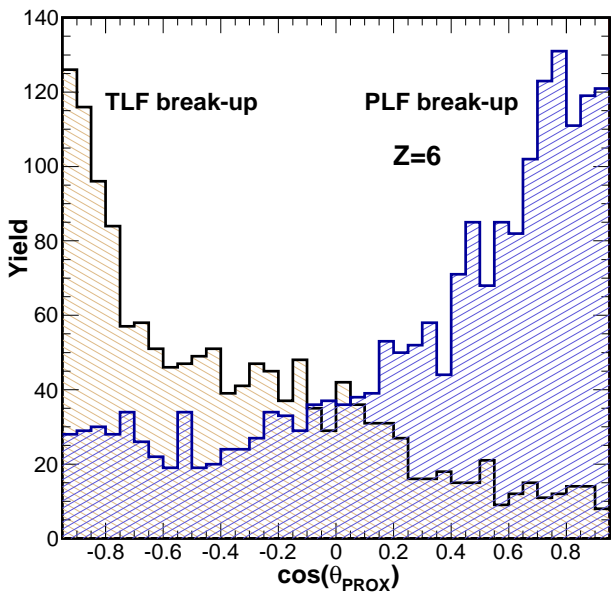}
\caption{Angular distribution $\cos({\theta}_{PROX})$ for IMFs of charge Z=6 with respect to the
PLF (blue hashed histogram) and with respect to the TLF (brown hashed histogram) for the direct kinematics $^{64}Ni$ + $^{124}Sn$ reaction. Aligned emission of the IMF towards midrapidity
corresponds to $\cos({\theta}_{PROX})\gtrsim$0.7 for PLF break-up and
$\cos({\theta}_{PROX})\lesssim$-0.7 for TLF break-up.
}
\label{figtimesca}
\end{figure}

These results were supported by different transport reaction simulations like stochastic
mean field (SMF) \cite{bar05,riz08} and constrained molecular dynamical model (CoMD-II)
\cite{pap07,pap01}.
The observed reaction dynamics is also consistent with kinematical simulations based
on pure three-body Coulomb trajectory calculations \cite{pia02}.
As an example, Figs. \ref{fig8qmd}a,b) shows a comparison for the same reaction of Fig. \ref{fig7}
with the constrained molecular dynamics (CoMD) model. We see that the experimental correlations close to the mid-rapidity IMFs velocities are well reproduced. The time for formation of these fragments is
estimated around 70 fm/\emph{c}, in good agreement with the framework of stochastic mean field calculations \cite{bar04}. Very interesting in Fig. \ref{fig8qmd}b), the CoMD simulation also include events where a PLF asymmetric fission is observed in a timescale of about 300 fm/c
(or longer).

A way to study the dynamical origin and alignments of mid-rapidity fragments is to consider the angle $\theta_{PROX}$ between the TLF and IMF-PLF center of mass direction $(\mathbf{n}_s= \mathbf{V}^{c.m.}_{PLF-IMF} - \mathbf{V}_{TLF})$
and the break-up axis defined by the relative velocity between the PLF and IMF oriented from the light to heavy system (see the inset in Fig. \ref{fig7}b)). Fig. \ref{fig7}b) displays the
$\cos({\theta}_{PROX})$ distribution for Z=8 and Z=18 charges. The distributions present a strong enhancement when $\cos({\theta}_{PROX})>$0.8, indicating a backward IMF emission with respect to the PLF with a strong alignment for $\cos({\theta}_{PROX})\approx 1$ along the PLF-TLF separation axis. The enhancement for the Z=18 charge is due to the onset of dynamical fission of the projectile \cite{rus10}.
A similar behavior has been shown recently for the PLF break-up in Ref. \cite{hud12} for the reactions $^{124}Xe+^{124,112}Sn$ at 50 \amev. By setting the condition $cos(\theta_{PROX})<$0 (forward emission in the PFL hemisphere) in Fig. \ref{fig7}b) we obtain the pattern of Fig. \ref{fig7}d), where the events populate the region along the axis r$_1$=1, as expected for a sequential decay of the IMF from the PLF source. In contrast, Fig. \ref{fig7}c) is obtained by selecting events with the conditions $\cos(\theta_{PROX})>$0.8 (backward emission), showing events located far from the $X$ and $Y$
axis and elongated orthogonally with respect the first diagonal, as expected for the dynamical emission of fragments.

\begin{figure*}[th]
\begin{center}
\includegraphics[width=0.75\textwidth]{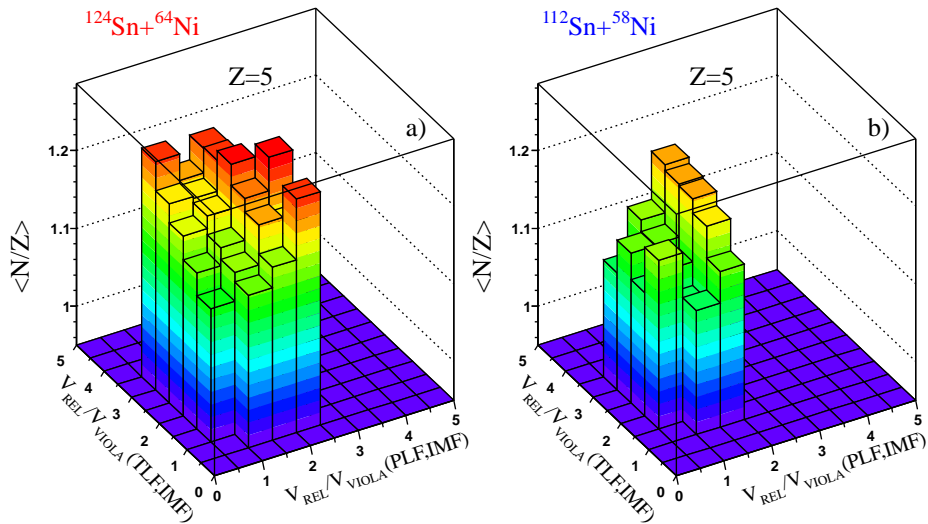}
\end{center}
\caption{For the $^{124}Sn$ + $^{64}Ni$, a) and $^{112}Sn$ + $^{58}Ni$, b)
reactions at 35 \amev\ and $M\leq 6$: tridimensional plot of
$<N/Z>$ for atomic number Z$_{IMF}$=5 and different bins in the
$r_1-r_2$ plane.
}
\label{figgratta}
\end{figure*}

The distribution of the $\cos({\theta}_{PROX})$ has been also studied in the direct
kinematics reactions $^{64}Ni$ + $^{124}Sn$ and $^{58}Ni$ + $^{112}Sn$ in order to complement the earlier study in reverse kinematics at the same beam incident energy \cite{def13}. This experiment was the first to use the CHIMERA silicon pulse shape upgrade, giving the possibility to identify in charge the
particles that were stopped in silicon detectors. In such a way, a unique set of information on
the mid-rapidity fragmentation process is available with the same apparatus, allowing the complete study of the competition between statistical evaporation and  mid-velocity emission, in the asymmetric system under study. Fig. \ref{figtimesca} shows, as an example, the angular $\cos({\theta}_{PROX})$ distribution for Z=6 IMF charge for the projectile-like (Ni-like, blue hashed histogram) and target-like (Sn-like, brown hashed histogram) break-up in ternary reactions for the $^{64}Ni$ + $^{124}Sn$ system. The separation between PLF emission or TLF emission is chosen gating on the relative velocities due to the mutual PLF-IMF or TLF-IMF Coulomb repulsion respectively. Conventionally the sign of $\cos(\theta_{PROX})$ for the TLF decay is taken in such a way that the peak at
$\cos(\theta_{PROX})\lesssim- 0.7$ represents the aligned break-up of Z=6 IMF towards midrapidity for TLF break-up, and following the previous definition, the peak at $\cos(\theta_{PROX})\gtrsim 0.7$ represents the aligned emission of IMF with charge Z=6 in the backward hemisphere with respect to the PLF,
for PLF break-up. The sum of the two contributions represents the cross section for
the dynamical IMF ternary emission at midrapidity.

We have in this way obtained an almost complete set of kinematical observables well correlated with
the time scale of the fragments formation. Starting from these analysis, the isotopic properties of IMFs have been studied for both dynamical and sequential emission, thus providing a powerful constraint
for transport theories.

\begin{figure}[b]
\includegraphics[width=0.50\textwidth]{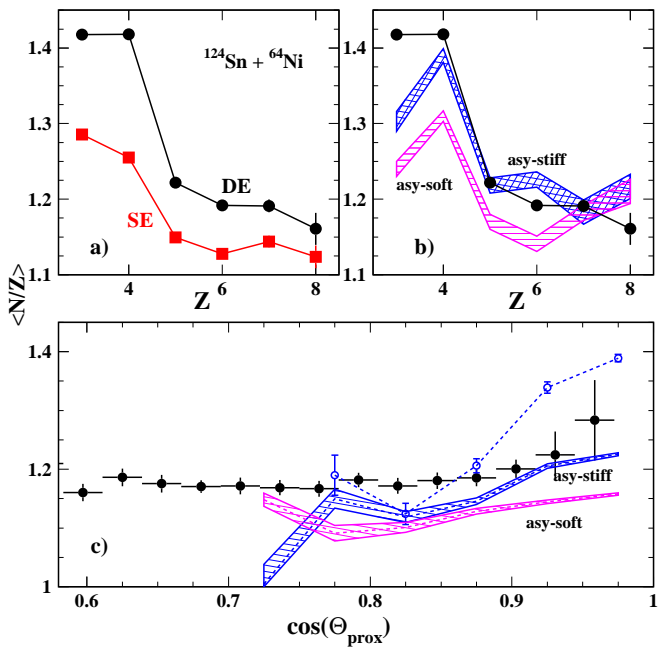}
\caption{For the $^{124}Sn+^{64}Ni$ reaction: a) experimental $<N/Z>$ distribution of IMFs as a
function of charge $Z$ for dynamically emitted (DE) particles (solid circles) and statistically emitted (SE) particles (solid squares); b) solid circles: same data as in panel a) for DE particles. Blue hatched area: SMF+GEMINI calculations for DE particles and asy-stiff parametrization; magenta hatched area: asy-soft parametrization; c) solid circles: experimental $<N/Z>$ as a function of $\cos(\theta_{PROX})$ for charges $5\le Z \le8$; empty circles: SMF calculations for primary fragments (asy-stiff parametrization); SMF+GEMINI calculations are indicated by blue hatched area (asy-stiff) and magenta
hatched area (asy-soft) respectively. The hatched zones indicate the error bars in the calculation. Adapted from Ref. \cite{def12}.}
\label{fig11}
\end{figure}

\subsection{Neck observables and constraints of the symmetry energy}
\label{sec:32}
A neutron enrichment of light charged particles (LCP) and IMFs in the midrapidity region in
comparison with the particles and fragments emitted by excited PLF
and TLF nuclei has been observed in experiments at medium heavy ion energies  \cite{koh11,pla08,pia06,luk97,she03,the06,bar13}.
In early studies, the dynamical character of the neutron enrichment
of the neck region was related by Sobotka \cite{sob94} to the isospin dependence of transport
processes in asymmetric nuclear matter, using the Boltzmann-Uehling-Uhlen\-beck (BUU) equation.

More precisely, theoretical simulations of heavy-ion dynamics have shown that the effective interactions modeling transport phenomena of neutrons and protons through the neck can be related to the slope (density gradient) and the value (isospin gradient) of the density dependent potential
term of the symmetry energy \cite{bar04,barP05,riz08}.

The symmetry energy, E$_{sym}$ includes a kinetic contribution with Pauli correlations and a potential contribution from the isospin dependence of the effective interactions. As useful parameterization of the potential part of symmetry energy as a function of the density, in modeling transport simulations, is given by a simple expression of the form:
\begin{equation}
E_{sym}= E_{sym}^{kin} + E_{sym}^{pot} = {C_{s,k}\over 2}(\rho/\rho_0)^{2/3} + {C_{s,p}\over 2}(\rho/\rho_0)^{\gamma}
\label{eqn1}
\end{equation}
In eq. \ref{eqn1} the parameter $\gamma$ determines the evolution of the symmetry energy
as a function of the density, where $C_{s,k}$ = 25 MeV and $C_{s,p}$ = 35.2 MeV are the constants for the kinetic and potential component, respectively \cite{tsa09}; thus different values of $\gamma$ indicate respectively \emph{asy-stiff} ($\gamma\ge 1$), where the potential part of symmetry energy increases with the density; or \emph{asy-soft} ($\gamma<1$), where it exhibits a flat behavior around the nuclear saturation density $\rho_0$ and it decreases at large density \cite{tsa12,bar05}.
The symmetry energy can be also described by means of its value $S_0$ at saturation density and by its first ($L$, slope) and second ($K_{sym}$, curvature) derivatives in a Taylor expansion around $\rho_0$ \cite{bar04}. The slope $L$ (or symmetry pressure) and $S_0$ constitute the key parameters in order to compare the symmetry energy probes from heavy-ion collisions to the astrophysics ones \cite{tsa12,lat12}.

Isospin transport phenomena are clearly strictly related to the
presence of a dilute region that is expected to be formed in the dynamical neck-like structure:
due to this density gradient, an excess of neutrons towards this dilute region (isospin migration) is predicted, depending upon the slope of the symmetry energy (that is larger for the \emph{asy-stiff} parametrization at sub-saturation density than for the \emph{asy-soft} one).
Conversely, if projectile and target have a large initial $N/Z$ asymmetry the isospin
transport drives the isospin equilibration trough the neck (isospin diffusion) which depends on the
magnitude of the symmetry energy at sub-saturation density; this latter is greater for the \emph{asy-soft} parametrization. As shown for example in \cite{lom10} the isospin migration and isospin diffusion both can compete to characterize the midrapidity and projectile/target residue isospin content.

In Ref. \cite{def12} we measured the degree of neutron enrichment for dynamically emitted IMFs (DE) at midrapidity for the reactions $^{124}Sn$ + $^{64}Ni$ and $^{112}Sn$ + $^{58}Ni$ at 35 \amev\ in comparison to the statistical emission (SE) from a PLF source. The selection was mainly based on fragments angular, velocity and isospin correlations as described in section \ref{sec:31} and more detailed in Ref. \cite{def12}.

To get more insights to the correlations between isospin, relative velocities, and emission
time-scale of IMFs we show in Figs. \ref{figgratta}a,b), for each bin (0.5 $\times$ 0.5 width)
in the plane $r_1-r_2$ of Fig. \ref{fig7}a), the average $N/Z$ isotopic distributions for Z$_{IMF}$=5
for both the neutron rich a) and the neutron poor b) systems. The largest values
of the neutron to proton ratios are obtained for events near
the diagonal of the $V_{REL}/V_{Viola}(PLF,IMF)$ vs. $V_{REL}/V_{Viola}\-(TLF,IMF)$
plane corresponding to prompt emission ($<$50 fm/c) and to the highest
degree of alignment.

Fig. \ref{fig11}a) shows the $<N/Z>$ as a function of the IMFs atomic number $Z$ for the neutron rich system $^{124}Sn$ + $^{64}Ni$. We observe that the value of $<N/Z>$ for the dynamically emitted particles (solid circles) is systematically larger with respect to the one for statistically emitted particles (solid squares). The same  effect can be seen for the neutron poor system (see Fig. \ref{fig21}).
In Fig. \ref{fig11}c) the correlations between the $\cos(\theta_{PROX})$ and $<N/Z>$ for all light fragments
with charge $5\le Z \le8$are reported (solid circles) for the same reaction.
An enhancement of $<N/Z>$ is observed at values of $\cos(\theta_{PROX})$ approaching to 1,
(maximum degree of alignment), corresponding to backward emission respect to the projectile. It is interesting to note that for the reactions $^{124}Xe+^{112,124}Sn$ at 50 \amev\ \cite{hud12} a similar enhancement at backward emission is also observed for light fragments, slightly dependent from the $N/Z$ initial content of the target nucleus ($^{112}Sn$ or $^{124}Sn$). Recently, the $N/Z$ of the emitted light fragments from PLF break-up was found clearly correlated with the neutron richness of the  target particularly when using an heavy neutron rich target, like in the reaction $^{64}Zn$ + $^{209}Bi$ \cite{bro13}. The authors deduced a rotation time from the alignment angle thus providing a nice
correlation between the enhancement of the $<N/Z>$ and the timescale for the lightest fragments
emitted in projectile break-up \cite{hud12,bro13}.

The experimental data have been compared with a transport model calculation in order to constraint
the density dependence of the symmetry energy. The stochastic mean field (SMF) model \cite{bar05},
based on Boltzmann-Landau-Vlasov (BNV) equation, was used. The SMF model implements the nuclear mean field dynamics as well as the effect of fluctuations induced by nucleon-nucleon collisions.
Two different parameterizations of the potential part of the symmetry term of EOS were used. The first one linearly increases with the density (asy-stiff) and the second one (asy-soft) exhibits a weak variation around the nuclear saturation density $\rho_0$. In the calculation shown
in Fig. \ref{fig11} and Ref. \cite{def12} the corresponding value of the slope parameter:
\begin{equation}
L=3\rho_0 \left({dE_{sym}(\rho) \over d\rho}\right)_{\rho=\rho_0}
\end{equation}
was found 80 MeV for the asy-stiff case and 25 MeV for the asy-soft one, respectively.
The value of S$_0$ (magnitude of the symmetry energy at $\rho_0$) is $\approx$30 MeV. The primary hot fragments are produced by SMF with an average excitation energy $E^*/A\approx2.5$ \amev\ and experience a de-excitation phase using the statistical model GEMINI \cite{char88}.

\begin{figure}[t]
\includegraphics[width=0.50\textwidth]{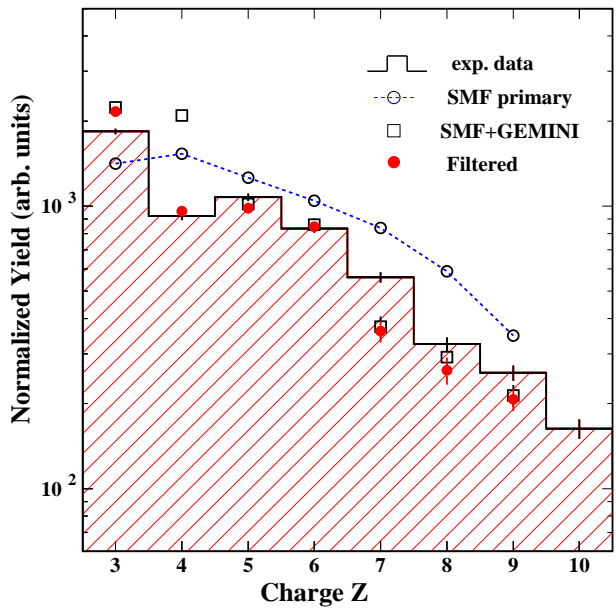}
\caption{Experimental charge distribution of dynamically emitted IMFs is compared, for the
$^{124}Sn$ + $^{64}Ni$ reaction, with SMF calculations (see text and Ref. \cite{def12} for details).}
\label{figsizesmf}
\end{figure}

In Fig. \ref{fig11}b,c) the SMF + GEMINI calculations are plotted as hatched area histograms for dynamically emitted fragments. Fig. \ref{fig11}b) shows the calculated $<N/Z>$ as a function of
the atomic number $Z$ of dynamical emitted IMFs. The two parameterizations for the symmetry energy term are indicated respectively by the blue hatched area for asy-stiff case and the magenta hatched area for the asy-soft one, including the statistical decay of the SMF primary fragments.
The asy-stiff parameterization, corresponding to a linear behavior of the potential part of symmetry energy with the density, produces more neutron rich fragments respect to the asy-soft choice and it
matches the experimental data fairly well. This behavior is also seen in Fig. \ref{fig11}c) where
the asy-stiff parameterization reproduces better than the asy-soft the $<N/Z>$ enhancement observed for the largest degree of alignment ($\cos(\theta_{PROX})>0.9$). For comparison, in Fig. \ref{fig11}c), the SMF calculations for primary fragments (asy-stiff) are also drawn (empty circles). In SMF calculations the primary neck emitted fragments display an exponential-like dependence with the charge \cite{bar04}. This is shown in Fig. \ref{figsizesmf} where the experimental charge distribution for dynamically emitted IMFs is compared with the SMF calculated distribution for primary fragments (empty circles), unfiltered final fragments (squares) and filtered by detector acceptance (filled circles).

\begin{figure*}[th]
\begin{center}
\includegraphics[width=0.70\textwidth]{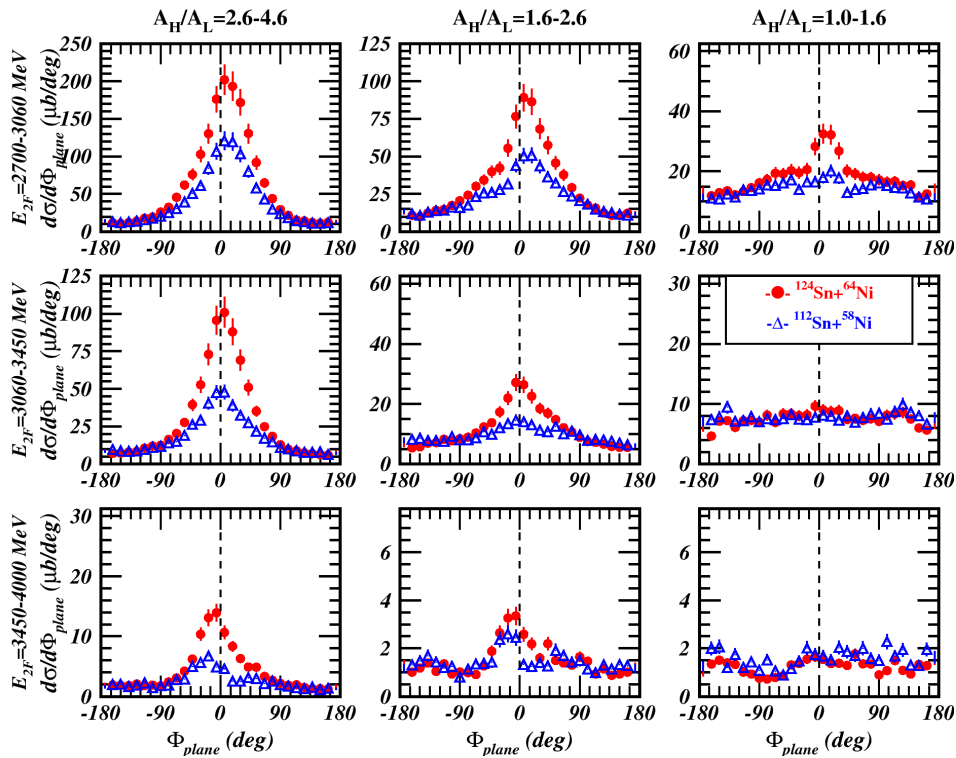}
\end{center}
\caption{``In-plane'' angular distribution of the PLF break-up fragments for the $^{124}Sn+^{64}Ni$ (circles) and $^{112}Sn+^{58}Ni$ (triangles) at 35 \amev, for three different bins of the total kinetic energy (E$_{2F}$) and mass asymmetry $A_H/A_L$. E$_{2F}$ is an indicator of the collision inelasticity (larger E$_{2F}$ values correspond to peripheral collisions). Adapted from Ref. \cite{rus10}.}
\label{fig13}
\end{figure*}

Thus, the correlation between the emission timescale of intermediate mass fragments (IMF) at mid-rapidity in semi-peripheral collisions and their isotopic composition gives strong indication that large values of $<N/Z>$ are acquired by light IMFs dynamically emitted in the early stage of the reaction. Comparison of these data with a SMF transport model points to a linear behavior of the potential symmetry energy around the saturation density. However, it is to be notice that quantitative evaluations
of the relevant parameters linking isospin transport phenomena with the symmetry energy density dependence are affected by relatively large error bars within the current experimental constraints of the symmetry energy, as obtained mainly by exploring isospin diffusion observables \cite{tsa12}.
Due to the importance to check the sensitivity to the symmetry energy
we are currently analyzing the fragments emitted in neck fragmentation using different observables for the reactions studied in both reverse and direct kinematics, and
different models comparison \cite{skw12}.

\section{From neck fragmentation to dynamical fission}
\label{sec:4}
With respect to the prompt neck emission, the emission of heavy IMFs (Z$>$8) from projectile-like
fragments splitting appears at a later stage with respect to the one observed for the light IMFs
neck fragmentation mechanism \cite{def05}. This observation is of crucial importance in studying the basic equilibration mechanism of the different degrees of freedom involved in heavy ion reactions, as a function of the energy and impact parameter. In fact, one is faced with emission of particles and
fragments that, starting from the early dynamical stage within a short time scale ($<$50 fm/c) evolves through a fast pre-equilibrium emission ($\approx$300 fm/c) until the typical  time scale of evaporation process is achieved.

Asymmetric binary splitting of projectile-like fragments  in two massive fragments has been interpreted as the break-up of a very elongated and deformed neck-like structure that drives the PLF towards a non equilibrated fast asymmetric fission during the interaction with the target. The works, at relatively low beam incident energy of Gl\"{a}ssel \emph{et al.} \cite{gla83} in $^{129}Xe$ + $^{122}Sn$ reaction at 12.5 \amev\ and Stefanini \emph{et al.} \cite{ste95}, in $^{100}Mo$ + $^{100}Mo$ and $^{120}Sn$ + $^{120}Sn$ collisions at 20 \amev, constitute the two main earlier studies on this subject. In those works three-bodies events in the final state of the reaction were mainly interpreted as a
relatively fast (around 300 fm/c) two-step sequential mechanism
of a fission-like splitting process following a deep inelastic collision. More recently, in experiments
at higher energies, two well defined PLF break-up components have been identified that contribute to IMF emission \cite{rus10,deff05,boc20,col03,mci10}: i) an aligned break-up with the recoil velocity of the projectile-like source suggesting a fast non-equilibrated fission process (dynamical fission); ii) an equilibrium (statistical) fission, where the angular distributions of the fission-like fragments are isotropic in the PLF reference frame. In this contest the time scale of the process as a function of
the incident energy and the reaction impact parameter can be considered the main signature among
different mechanisms: from early neck prompt fragmentation (40$-$120 fm/c) towards dynamical fission
(120$-$300 fm/c) \cite{deff05,ste95,mci10,cas93} to equilibrated fission that is at least
10$-$100 times longer.

However, by increasing  the coulomb field of the colliding heavy  systems new phenomena could be expected. In fact, it has been recently reported, in the frame of CHIMERA collaboration \cite{wil10,skw08}, the presence of a new mechanism in the collisions $^{197}Au$ + $^{197}Au$ at
15 \amev. It has been shown that, in inelastic collisions, the PLF/TLF nuclei break into three or four massive fragments of comparable size with nearly aligned configuration along a common re-separation axis. A surprising fast time scale for the process ($<$100 fm/c) has been estimated. This break-up mechanism shows the characteristics of a deeply inelastic collisions, but it occurs at smaller impact parameters with respect to what is expected in the well known low-energy dissipative binary reactions \cite{wil110}. The new fragment production is different from both the asymmetric massive break-up observed at Fermi energies and the neck fragmentation process involving light IMFs \cite{wil10,li13}.

The main experimental signature of dynamical asymmetric fission in two massive fragments
is that the heaviest of the two fission fragments (usually the fastest one)
is forward directed, while the lightest fragment is emitted preferentially between its heavy partner
and the target-like nucleus, thus resulting in a aligned three body configuration.
In contrast, in the equilibrated statistical fission the PLFs angular distribution
is forward-backward symmetric  because the emission is isotropic in the reference frame of
the fissioning PLF. In the Ref. \cite{deff05} relative to the reaction $^{124}Sn$ + $^{64}Ni$ at 35 \amev\ the onset of dynamical fission process was clearly well characterized: the nonequilibrium properties of the fragments undergoing a dynamical fission process were observed in both in-plane and
out-of-plane angular distributions, as well as in the relative velocities of the two fragments and
their alignment properties (see below). However, in contrast to light fragments originating
from the dynamical neck fragmentation, heavy fragments ($Z>8$) show velocities approaching to the projectile one and their corresponding invariant cross sections (in the V$_\parallel$,V$_\perp$
plane) indicate clear signature of the persistence of Coulomb ring structures around the PLF velocity.
The time scale for dynamical emission was estimated in a relatively short time window after collision, i.e., $100 < t < 300$ fm/c.

\begin{figure}[bh]
\includegraphics[width=0.50\textwidth]{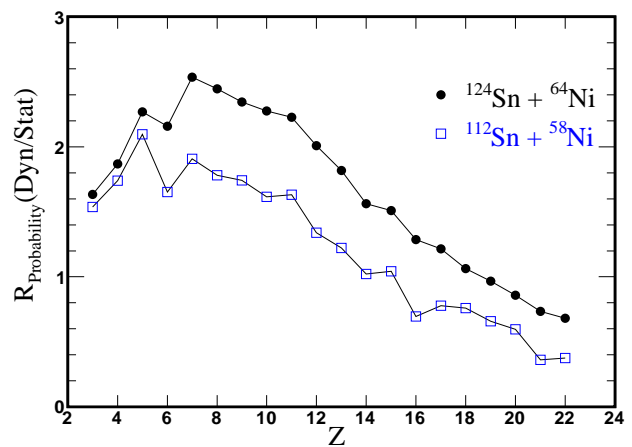}
\caption{Ratio between probability of dynamical and statistical fission as a function of the
IMFs charge for the $^{124}Sn+^{64}Ni$ (circles) and $^{112}Sn+^{58}Ni$ (squares) reactions.}
\label{fig14}
\end{figure}

Dynamical fission was found very sensitive to entrance channel isospin. In fact, the most striking result in the onset of dynamical fission studied with the CHIMERA detector was found in measuring surprising different values of the production cross section in two reactions representing a neutron rich $^{124}Sn$ + $^{64}Ni$ and a neutron poor system $^{112}Sn$ + $^{58}Ni$ \cite{rus10}, and thus
providing evidence for the first time of the isospin dependence in this mechanism.
The measurement and the event selection method are described in Ref. \cite{rus10,deff05}.
Very briefly, semi-peripheral events were selected by requiring values of the impact parameters
larger than $b_{red} = b/b_{max}\ge0.7$ for both systems, and showing at least
two fragments (H,L) forward emitted (in the center-of-mass system) with a total charge
$Z_{2F}=Z_{H}+Z_{L}$ (H and L stand, respectively for the heavy and the light fragment)
close to the charge of the PLF (Z=50).
Fig. \ref{fig13} shows the differential angular cross-section for the two systems for three bins of the mass asymmetry $A_H/A_L$ (columns) in the investigated range $1\leq A_H/A_L<4.6$ and for
three ranges of the total kinetic energy $E_{2F}=E_H+E_L$ of the two fragments (rows) as a function of the ``in-plane'' $\Phi_{plane}$ angle. $\Phi_{plane}$ is the PLF-fission angle projected in the reaction plane. The value of $\Phi_{plane}=0$ corresponds to the heaviest of the two fragments (H) moving forward along the PLF flight direction while the lightest fragment (L) is emitted backward along the PLF-TLF direction. In Fig. \ref{fig13} we observe, with increasing the mass asymmetry and inelasticity (lower E$_{2F}$), a rise of the forward peaked component superimposed to a flat $\Phi_{plane}$ distribution, this last is characteristic of equilibrium fission. This ``equilibrium'' component is predominant for peripheral collisions and its cross-section is approximately the same for the two systems. On the contrary the estimate of the cross-sections for the dynamical and equilibrated fission components indicates that the dynamical component is larger for the neutron rich system by a factor of about 2 as compared with the neutron poor system \cite{rus10}.

\begin{figure}[b]
\includegraphics[width=0.50\textwidth]{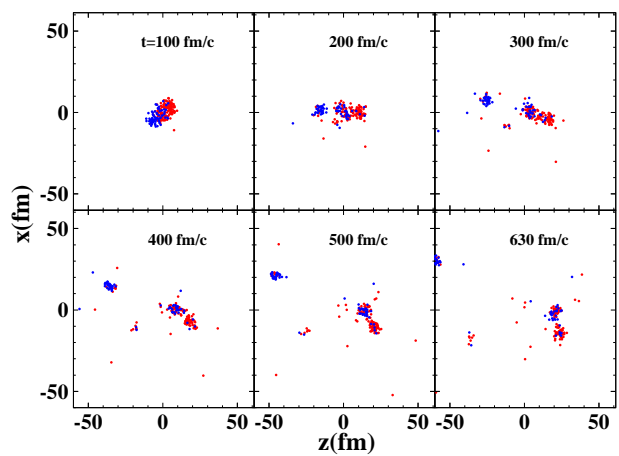}
\caption{Snapshot at different time steps of one event related to the $^{124}Sn+^{64}Ni$
system at 35 \amev\ leading to the fission of the heavy partner. CoMD-II simulation
adapted from Ref. \cite{pap07}.}
\label{figfiss}
\end{figure}

This surprising result can be observed also in Fig. \ref{fig14}  where the obtained ratios of dynamical and statistical contributions are reported as a function of the IMFs charge \cite{prus13}. Dynamical emission is favored for the neutron rich systems. Conversely a similar effect has not been observed when comparing the binary break-up of the projectile in the two reactions $^{124}Xe$ + $^{112}Sn$ and $^{124}Xe$ + $^{124}Sn$ at 50 \amev\ \cite{hud12} studied with the FIRST array \cite{mci10}. In these reactions only the $N/Z$ of the target was varied, indicating no dependence upon the $N/Z$ of the target on the PLF dynamical break-up process. Indeed the different condition of this experiment with respect
to the one investigated by CHIMERA renews the interest about the dependence of the isospin dynamics as a function of the isospin observable and the energy.
Although the origin of the strong enhancement in the neutron rich system of the dynamical
component with respect to statistical observed at 35 A.MeV\ appears to be related to the initial isospin $N/Z$ of the system, a clear interpretation of this result still remains an open question from both experimental and theoretical sides and further  investigations are needed. Recently a new experiment has been performed with the CHIMERA detector by studying the reaction $^{124}Xe$ + $^{64}Zn$ at 35 \amev, with the intention to investigate a neutron poor system, but having the same projectile
and target mass of the neutron rich $^{124}Sn$ + $^{64}Ni$ system \cite{rusexp}, in order
constraining the isospin dynamics through the comparison of these two isobaric systems.

From theoretical side, the dynamical fission phenomena have been studied by transport model
simulations. The stochastic Boltzmann-Nordheim-Vlaslov simulation (BNV) \cite{bar04} is a good
approach in order to describe the gross features of dynamical fission phenomena.
However, the simulation can hardly follow the complete time evolution of the deformed PLF up to the scission point. Indeed, quantum molecular dynamics models (QMD), like the Constrained Molecular Dynamics (CoMD-II) or the improved quantum molecular dynamics (ImQMD) have shown that can reproduce
many features of the dynamical fission processes, respectively in the field of the
asymmetric break-up of PLF(TLF) in the Fermi energy domain \cite{pap07} and in the
description of fast ternary break-up of heavy fragments at lower energies \cite{tia10}. Fig. \ref{figfiss} shows the time evolution on a long
timescale of a typical event leading to a PLF fission in the framework of a CoMD-II calculation \cite{pap07}.
The main feature of this last model is a self-consistent \emph{N}-body approach that solves the
equations of motion using procedures to satisfy event-by-event the Pauli principle and
the total angular momentum conservation law.

\begin{figure}[b]
\includegraphics[width=0.50\textwidth]{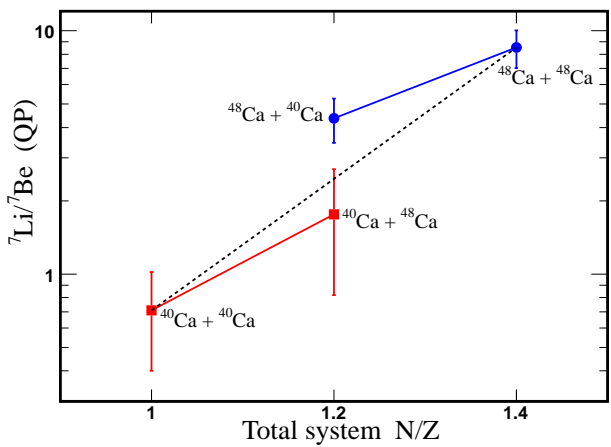}
\caption{Experimental projectile-like isobaric ratios $^{7}Li/^{7}Be$ in the reactions $^{40}Ca + ^{40,48}Ca$ and $^{48}Ca + ^{40,48}Ca$. The dotted line (``equilibration line'') considers
the linear relationship between $ln(^7Li/^7Be)$ and the $N/Z$ of the source.
Adapted from Ref. \cite{lom13}.}
\label{fig16}
\end{figure}

\section{Isospin diffusion and stopping in Chimera data}
\label{sec:5}
One of the most prominent features of heavy ion collisions at medium energy is the unique
possibility to probe the effective symmetry potential of EOS at sub-saturation densities
by studying characteristic phenomena related to the isospin dynamics which are due to the density
gradients in strongly interacting systems and isospin asymmetries, such as isospin migration
and nucleons diffusion. Interesting analysis have been employed, with different degree of
sensitivity, such as, isospin diffusion observable \cite{tsa09,tsa04}, double ratios
involving neutron and proton energy spectra \cite{fam06} and transverse collective flows of light
charged particles \cite{koh11}. These analysis involve delicate comparison of the experimental data
with time consuming dynamical simulations by including sequential de-excitation decay of the primary fragments. Briefly, as seen in sect. \ref{sec:3}, a migration of neutrons and protons shaped by the baryonic density and isospin diffusion are predicted by transport simulations. At medium energy, in binary reactions induced by projectile and target nuclei with different $N/Z$ asymmetry a substantial exchange of nucleons (isospin diffusion) between  projectile-like and target-like nuclei takes place driving the binary system toward an uniform isospin $(N-Z)/A$ distribution (isospin equilibration).

In reactions at 50 \amev\ induced by Sn isotopes \cite{tsa04} the time for re-separation of the PLF and
TLF was found around 100 fm/c, but only a partial equilibrium of the isospin asymmetry was deduced.
The degree of isospin equilibration attained depends upon many factors, such as the interaction time,
the impact parameter, the incident energy and it is affected by pre-equilibrium emission, Coulomb
effects and secondary decays.
Transport models predict that the isospin diffusion depends linearly on the absolute value of the symmetry energy at subsaturation density \cite{barP05,riz08} thus providing important information on transport properties of asymmetric nuclear matter. Because at sub-saturation densities a soft ISO-EOS parameterization shows a larger value compared to the stiff one \cite{riz08}, it is expected that the soft behavior of the potential part of the symmetry energy mainly determines the isospin equilibration.

The study of isospin diffusion as a function of dissipation (or centrality) in the collision
performed by the INDRA collaboration \cite{gal09_1} in projectile fragmentation
for two beams energies (52 and 74 \amev) confirms this scenario and
show a clear evolution of the isospin dynamics from transparency to equilibration
with the increase of the dissipated energy. At 52 \amev\ a better overall agreement with experimental
data is obtained for the asy-stiff case (symmetry term linearly increasing with nuclear density). However, at the highest incident energy, 74 \amev, the same data indicate that these effects
are almost independent by the symmetry energy parameterization (asy-soft or asy-stiff)
chosen to describe the data.

In the following, the results of two measurements performed with the CHIMERA array for the reactions
$^{40}Ca$ + $^{40,48}Ca$ and $^{48}Ca$ + $^{40,48}Ca$ at 25 \amev\ \cite{lom10,lom13} and $^{112}Sn$ + $^{112,124}Sn$, $^{124}Sn$ + $^{112,124}Sn$ at 35 \amev\ \cite{sun10} are briefly discussed.

\begin{figure}[t]
\includegraphics[width=0.50\textwidth]{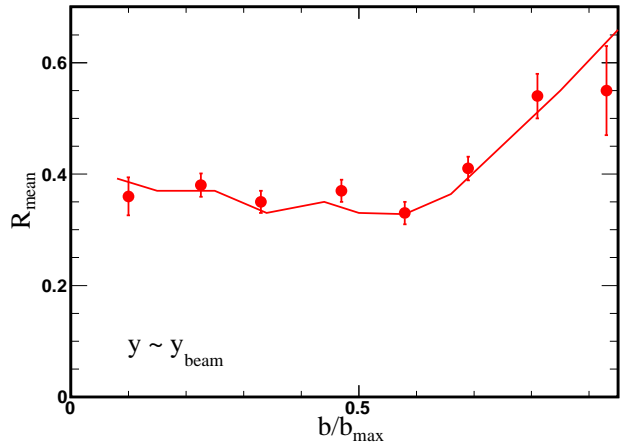}
\caption{Isospin transport ratio $R_{mean}$ as a function of reduced impact parameter constructed using $^7$Li and $^7$Be fragments near projectile rapidity. Ratios measured in rapidity regions symmetric with respect to mid-rapidity have been combined together in the mean value $R_{mean}$. The line is the result of ImQMD calculations (see text). Adapted from Ref. \cite{sun10}.}
\label{fig17}
\end{figure}

The system $^{40}Ca$ + $^{48}Ca$ reactions at 25 \amev\ has been investigated for  semi-peripheral collisions (impact parameter b=7 fm)  by three different models: Molecular Dynamics \cite{pap01}, and  two transport simulations based, respectively,  on Boltzmann-Nordheim-Vlasov (BNV) \cite{bar05} and Boltzmann-Uehling-Uhlenbeck (IBUU04) \cite{Li05}. Qualitatively similar results have been found in
these models by looking at the evolution of the $N/Z$ of projectile-like and target-like sources. In particular, at the re-separation time of about 150 fm/c, calculations show that the complete charge equilibration is not yet reached \cite{lom10}.

\begin{figure}[th]
\includegraphics[width=0.50\textwidth]{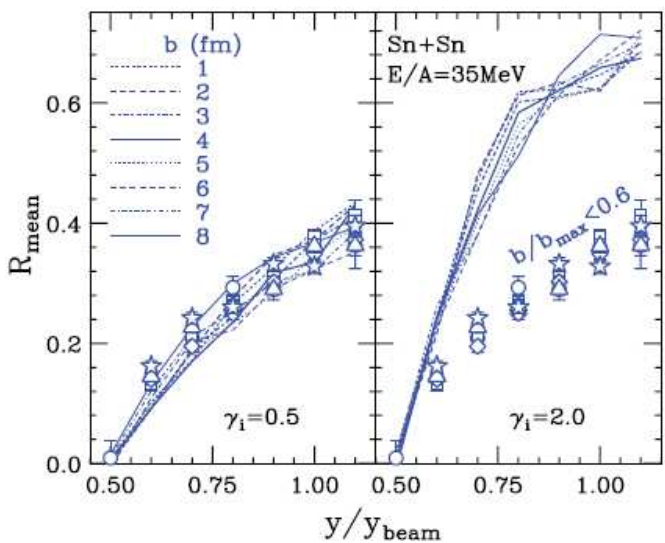}
\caption{Isospin transport ratio $R_{mean}$ as a function of rapidity normalized to the beam rapidity. Left: data are compared with ImQMD calculations with $\gamma=0.5$. Right: the same data are compared with a stiffer symmetry potential with $\gamma=2.0$. The symbols correspond to different reduced impact parameters $b/b_{max}$ equal to respectively: 0.1 (open circle), 0.23 (open square), 0.34 (open diamond), 0.46 (open triangle) and 0.58 (open star) in the data. Adapted from Ref. \cite{sun10}.}
\label{fig18}
\end{figure}

In order to study the $N/Z$ equilibration of the projectile-like source (PLF) the yield ratios of the pair of isobars $^7Li/^7Be$ was used. This isobaric ratio $ln(^7Li/^7Be)$ is expected to be linearly correlated to the asymmetry of the source \cite{liu07}.
Fig. \ref{fig16} shows as a function of the $N/Z$ of the total system, the isobaric ratios for the PLF
source for the four studied systems. For the symmetric systems ($^{40}Ca$ + $^{40}Ca$ and $^{48}Ca$ + $^{48}Ca$) no isospin diffusion is expected, thus these systems can be used as good reference.

For the mixed system a net neutron exchange from neutron rich towards the neutron poor system
is observed, but two different values of the isobaric ratios are found with clear signature of a memory effect of the $N/Z$ of the projectile.
Evidently, at the investigated energy, the diffusion path of the isobaric ration of PLF from the neutron poor $^{40}Ca$ + $^{40}Ca$ to the neutron rich $^{48}Ca$ + $^{48}Ca$ is strongly influenced by isospin content $(N/Z)_P$ of the projectile.

The dotted line in Fig. \ref{fig16} takes into account the linear relationship between $ln(^7Li/^7Be)$ and the $N/Z$ of the source for symmetric systems. The isobaric ratio $^7Li/^7Be$
is used to estimate the average $N/Z$ of the projectile-like $(N/Z)_{QP}$. Following formalism
described in Ref. \cite{lom13,kek10} the fraction $f_{EQ}$ of $N/Z$ equilibration is given by:
\begin{equation}
f_{EQ} = {{\left(N \over Z\right)_{QP} - \left(N \over Z\right)_{P}}
\over {\left(N \over Z\right)_{TOT} - \left(N \over Z\right)_{P}}}
\end{equation}
where $(N/Z)_{TOT}$ is the $N/Z$ of the Ca+Ca composite system.
A partial degree of equilibration ($f_{EQ}\approx 0.63$) has been evaluated ($f_{EQ}=1$ for complete equilibration), very similar to the results reported in Ref. \cite{kek10} for $^{40,48}Ca$ +$^{112,124}Sn$ at 32 \amev. In the same context and for large energy dissipation, data from INDRA collaboration \cite{gal09_1} show that isospin equilibration can be reached at incident energy of 52 A MeV, for the Ni+Au system.

The isospin diffusion has been studied at MSU-NSCL for the reactions $^{112}Sn+^{112}Sn$,  $^{112}Sn+^{124}Sn$, $^{124}Sn+^{112}Sn$, $^{124}Sn+^{124}Sn$ at 50 \amev\ \cite{tsa09,tsa04,liu07}
using the LASSA array \cite{dav01}. The same reactions have been studied at 35 \amev\ incident energy at LNS using the CHIMERA detector by MSU-CHIMERA collaboration with the main goal to extend the isospin diffusion studies at lower energies where longer interaction times take place with respect to the same reaction at 50 \amev, and consequently smaller $N/Z$ transparency effects are expected.
In fact, the results of experiments at 50 \amev\ showed that the reaction time is not long enough to lead to a complete $N/Z$ equilibration \cite{tsa04}. The isospin diffusion between Sn projectiles and target nuclei has been studied in both experiments by inspecting the isospin transport imbalance ratio $R_i$ \cite{ram00}:
\begin{equation}
R_i = {2x_i -x_{A+A}-x_{B+B}\over x_{A+A} - x_{B+B} }
\end{equation}
where $x$ is an isospin sensitive observable, ``A+A'' refers to the neutron poor symmetric system ($^{112}Sn+^{112}Sn$) and ``B+B'' to the neutron rich one (see Ref. \cite{liu07} for more details).

In the above formula the observable $x$ was the isobaric yield ratio $ln(^7Li/^7Be)$. The effects of isospin diffusion has been studied looking at the evolution of $x$ from midrapidity to the PLF rapidity at different impact parameters. Imbalance ratios measured in rapidity regions symmetric with respect to mid-rapidity have been combined together, so obtaining  the mean value $R_{mean}$.
Fig. \ref{fig17} shows, as a function of the reduced impact parameter the $R_{mean}$ value
evaluated for $^7$Li and $^7$Be fragments near projectile rapidity. For imbalance ratio $R=0$ in the projectile rapidity region, a complete equilibration is expected. Conversely $R$ near to unity
indicates a complete transparency (no isospin equilibration).
$R_{mean}$ reaches a saturation value for $b/b_{max}<0.6$ indicating that a large degree of isospin transparency (or incomplete stopping) persists in the reaction.
These results are consistent with the extensive studies of stopping in central Xe+Sn collisions at Fermi energies performed with the INDRA 4$\pi$ detector \cite{leh10}.

In order to constraint  the density dependence of the symmetry energy the data taken at 35 \amev\
with CHI\-MERA detector have been compared with calculations performed with the Improved Quantum Molecular Dyna\-mics (ImQMD) \cite{zha05}. This model provides consistent results when compared to data from different observables like isospin diffusion and ratios of neutrons and protons spectra \cite{tsa09,zha12}.
In this model the potential contribution to the symmetry energy (Asy-EOS)
is parameterized as $S(\rho)=C_{s,p}/2\times(\rho/\rho_0)^\gamma$ where $C_{s,p}=35.2$ MeV
(see Eq. \ref{eqn1})
and $\gamma$ is the parameter shaping the stiffness of the symmetry energy. Fig. \ref{fig18} shows the isospin transport ratio $R_{mean}$ as a function of the normalized rapidity for different impact parameters $b/b_{max}<0.6$.

The data are compared with ImQMD model (lines) by adopting  two parameterizations for
the density dependence of Asy-EOS, a soft behavior with value $\gamma=0.5$ (left panel)
and stiff with $\gamma=2$ (right panel). The simulations in the left panel show that the isospin transport ratios were better reproduced by a softer term of the symmetry energy.
The same parameterization
was used to describe the data of Fig. \ref{fig17}. For this calculation,  the saturation density
of value $S(\rho_0)=30.5$ MeV and the slope parameter  of value $L\approx 50$ MeV were extracted,
as shown by a star in Fig. 2 of Ref. \cite{tsa12}, where  the present ``status of the art''
of symmetry energy evaluation at sub-saturation density, as obtained in heavy ion collisions
were compared with other probes. It as to be noticed, looking at this plot, that within experimental uncertainties, density dependence of symmetry energy as obtained in terrestrial experiment
shows a poor overlap with data from astrophysical observations \cite{ste12}.
However, recent new constraints on the symmetry term of EOS from neutron star mass-radius relation, combining different models estimate for $L$ a range of values between 41 and 84 MeV to
within 95\% confidence \cite{ste13}, that is consistent with the current constraint obtained at low densities from heavy ion reactions and other probes from nuclear structure studies \cite{car10,roca11}.
Thus more precise constraints in the Fermi energy domain reducing the present uncertainties
and with the necessity of new experimental data at super-saturation densities represent a
real challenge for the next future.

\section{Isoscaling. Odd-Even effects in Z and N distributions}
In this section are discussed odd-even effects of proton and neutron numbers
in the cross section (staggering) of fragments produced by different mechanisms in heavy ion
collisions and in the neutron to proton ratio $N/Z$.
Also, a brief description of such studies with the CHIMERA array is reported.
Production cross section distributions (isotopes, isotones, masses) of IMF and light charged
particles (and yield ratios of fragments) have been used as basic observables to constraint
the symmetry energy. Evidently, from an experimental point of view, such studies
require experiments allowing for both charge and mass identifications
in a broad range of atomic numbers \cite{carb12}.

\begin{figure}[th]
\includegraphics[width=0.50\textwidth]{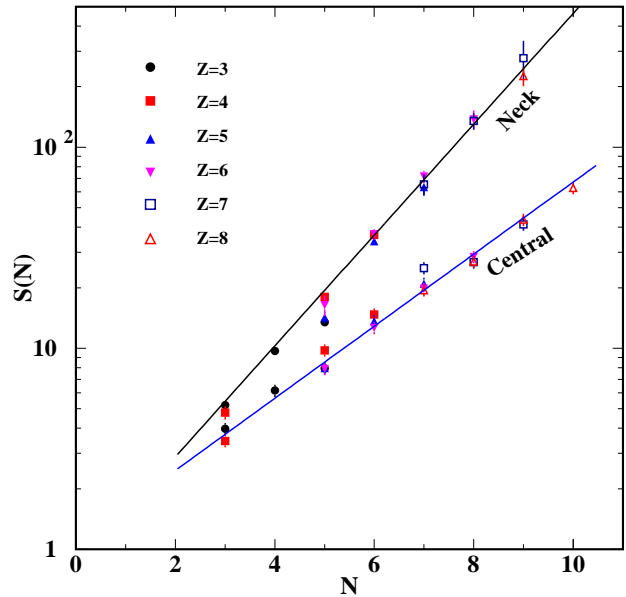}
\caption{Scaled isotopic ratio plotted as a function of the neutron number for central
collisions and neck fragmentation emission in Sn+Ni reactions at 35 \amev. Adapted from Ref. \cite{def06}.}
\label{figisosca}
\end{figure}

An isotopic scaling law (isoscaling) in multifragmentation or in the projectile fragmentation
was deduced \cite{xu00,ger04,tsa01,lef05,col06,gal10}. This scaling law is described in the relation:
\begin{equation}
R_{12}(N,Z)=Y_2(N,Z)/Y_1(N,Z) = C\exp(\alpha N+\beta Z)
\label{iso}
\end{equation}
or by the equivalent scaled ratio:
\begin{equation}
S(N) = R_{12}\exp(-\beta Z) \simeq C\exp(\alpha N)
\end{equation}
where $\alpha$ and $\beta$ are the scaling parameters, C is a normalization constant
and the index 2 and 1 correspond respectively to the yield (cross section) for a neutron rich and neutron poor reactions at the same bombarding energy.

Scaling law behavior have been predicted for both statistical multifragmentation \cite{tsa01,bot02} and dynamical models  where the parameter $\alpha$ has been linked to symmetry energy and the apparent temperature of emitting source \cite{she07,ono03} as it follows:
\begin{equation}
\alpha={4E_{sym} \over T} \left[ {Z^2_1 \over A^2_1}  - {Z^2_2 \over A^2_2 } \right]
\label{iso1}
\end{equation}
where, $Z_1/A_1$ and $Z_2/A_2$ are (in statistical models) the charge and the mass numbers of the two initial equilibrated fragmenting systems used in the numerator and denominator of eq. \ref{iso} and $T$ is the effective temperature, supposed to be the same for the two systems.
In fact, isoscaling behavior does not imply that a full thermodynamical equilibrium is reached before fragment emission \cite{col06,dor11,col05}.

A decrease of the symmetry energy with the increasing of the excitation energy and centrality
of the reaction has been observed in the framework of statistical models
(see \cite{mar12} and references therein).
In the reactions $^{124,112}Sn + ^{64,58}Ni$ at 35 \amev\ studied with the CHIMERA detector
isoscaling was observed in light fragments $Z\le8$ for the most central collisions, as selected by a multidimensional tensor analysis method \cite{ger04} and the value $\alpha=0.44\pm0.01$ was evaluated.
Data for this reaction are reported in Fig. \ref{figisosca} where the scaled ratio S(N) is drawn as a function of fragments $N$ number. The estimation of the symmetry energy for central collisions was
around 12 MeV (with an uncertainty of at least 30\%) \cite{ger07} and confirmed the general trend, observed in other experiments, of a decrease of the symmetry energy with centrality
of the collisions and excitation energy of the emitting source.
Interesting, in the same Fig. \ref{figisosca} it can be seen a scaling law,
as extracted for neck fragmentation where a steeper value of parameter $\alpha=0.62\pm0.02$
was found \cite{def06}. Neck fragmentation is a phenomenon unambiguously related to fast dynamical process (see section \ref{sec:3}).
Indeed an isoscaling behavior for the primary neck fragments is reported in stochastic mean field simulations \cite{bar04}. The steeper dependence of the $\alpha$ parameter for neck fragments
(with respect to the one observed for central collisions) indicates a neutron enrichment of the neck fragments by neutron migration from PLF(TLF) toward the low density region of the neck. The persistence of the neutron enrichment after secondary decay is consistent with a moderate average
excitation energies (as deduced by the light particle observed multiplicity) of the neck fragments.
Some studies on isoscaling analysis for fragments emitted by projectile break-up in dissipative reactions  (CHIMERA data) for the systems $^{48}Ca$ + $^{124,112}Sn$ at 45 \amev\ are in progress \cite{sin13}.

\begin{figure}[bh]
\includegraphics[width=0.50\textwidth]{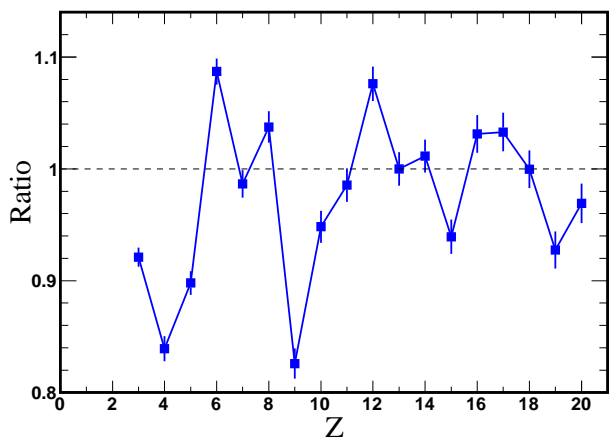}
\caption{Ratio of charge distributions between neutron poor $^{112}Sn+^{58}Ni$ and
neutron rich $^{124}Sn+^{64}Ni$ $Y(Z)$ yields measured in central collisions. Adapted form
Ref. \cite{ger04}.}
\label{fig20}
\end{figure}

When using the isoscaling observable to study the symmetry energy as a function of the temperature
and reaction mechanism much care has to be  devoted to  minimize or to evaluate the quantitative distortions induced in the primary fragment cross sections due to pre-equilibrium emissions,
secondary decays, staggering and feeding from unstable states \cite{col05,mar12,wue09}.

A characteristic odd-even effect in charge (mass) distribution is given by a staggering behavior of the charge (mass) production cross section of the detected fragments. A staggering  effect has been observed in heavy ion collisions in a broad range of incident energies (from 5 to 1000 A MeV) and different reaction mechanisms \cite{ric04,dag12,lan92,yan99,win00}.
In fact, staggering is found in low energy heavy ions physics, in particular in the compound nucleus and its decaying modes of excited residues \cite{ade11}. With the progressive increase of the incident energy it was observed in deep inelastic and projectile fragmentation reactions at the Fermi
energies up to the relativistic regime. Due to this large energy range, interpretations about odd-even effects and their relation with the isospin dependence of the reaction mechanism
are still controversial.

From one side staggering effects have been interpreted as determined by nuclear structure
properties (mainly link\-ed to nuclear pairing forces) at the end of the evaporation cascade
\cite{ric04}; on the other side it was suggested that the reaction mechanism for
fragments formation plays a substantial role \cite{dag11}.

\begin{figure}[bh]
\includegraphics[width=0.50\textwidth]{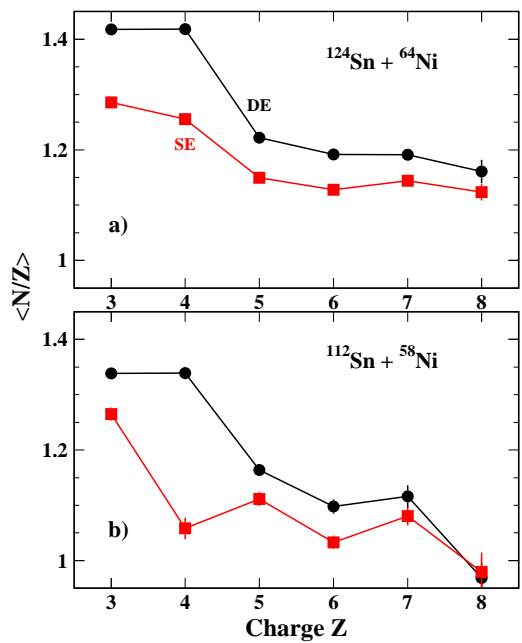}
\caption{Experimental $<N/Z>$ distribution of IMFs
as a function of charge Z for statistical emitted particles (solid
squares) and dynamical emitted particles (solid circles), for the
reactions: a) $^{124}Sn+^{64}Ni$ and b) $^{112}Sn+^{54}Ni$.
Adapted from Ref. \cite{def12}.}
\label{fig21}
\end{figure}

Efforts have  been devoted to link staggering effects with isospin physics,
and in particular, with the symmetry energy, that is mainly the object of this review.
A first link, by quantum molecular dynamics calculations, was the hypothesis that the strength of the symmetry energy had a direct effect in the amplitude of charge distribution
of fragments \cite{su11}. A second one was mainly related to the study of the temperature
dependence of $E_{sym}$ below the saturation density, in central collision
multifragmentation reactions. For example, it has been suggested from statistical microcanonical calculation that second momentum (widths) of the isotopic distributions as well as the
isoscaling parameters could give a direct estimation of $E_{sym}$ \cite{rad07}.
However, in this evaluation it is of crucial importance to take into account
the secondary decay in order to deduce the primary partitions of the nuclear ensemble at freeze-out from the detected secondary cold fragments. The influence of secondary decay on
staggering \cite{dag12,wink13} is one of the suggested method to trace back this information
along the evaporation chain.

In Fig. \ref{fig20} the ratio between the fragment charge distributions measured in central collisions for $^{112}Sn$ + $^{58}Ni$ and $^{124}Sn$ +$^{64}Ni$ reactions at 35 \amev\ \cite{ger04} is shown.
An enhancement in the production of even-to-odd atomic number $Z$ of the fragments is
clearly evidenced. In Fig. \ref{fig20}, the staggering effect is emphasized by looking at the ratio between the charge distribution of the neutron-poor reaction and the neutron-rich one.
Another interesting effect evidenced for the same reaction is shown in Fig. \ref{fig21} \cite{def12}, where (see Sec. \ref{sec:32}) the experimental $<N/Z>$ distribution of IMFs as a function of
atomic number Z for both statistical (sequential decay) and dynamical emitted particles in semi-peripheral collisions for the neutron poor reaction $^{112}Sn$ + $^{58}Ni$ and
for the neutron rich one $^{124}Sn$ + $^{64}Ni$ are plotted. It is clearly observed a flattening
of the odd-even effect in the $<N/Z>$ distribution of the neutron rich system with respect to the neutron poor one. It is important to notice that the amplitude of the odd-even effect is enhanced (clearly seen in Fig. \ref{fig21}, for the neutron poor system) for fragments dynamically emitted by the neck (DE) with respect to the ones statistically emitted in sequential decay (SE), so, proving the crucial role of the reaction mechanism.

Staggering effects have been associated with the nuclear pairing forces that link the one-proton
(one-neutron) separations energies for even Z (or N) distributions of the fragments in the last steps of the decay chains with the nuclear binding energies \cite{win00}. In fact one-proton separation energies for even-Z are higher than the one for odd-Z so making nuclei with even proton number more stable than the ones with odd proton number \cite{frie07}. Enhancement of the $N/Z$ ratio is so linked to the higher yields of even-Z nuclei with respect to the odd-Z ones, as shown in Fig. \ref{fig20}.

Anyway the recent work of Ref. \cite{dag12} has shown that the staggering effects are present not only in the last step of the decaying chain but also in the decay from excited levels or from low-lying resonances, thus asserting the importance to look at the whole decay chain of
the de-exci\-tation process and to the time scale of the production mechanism.

\begin{figure}[b]
\includegraphics[width=0.50\textwidth]{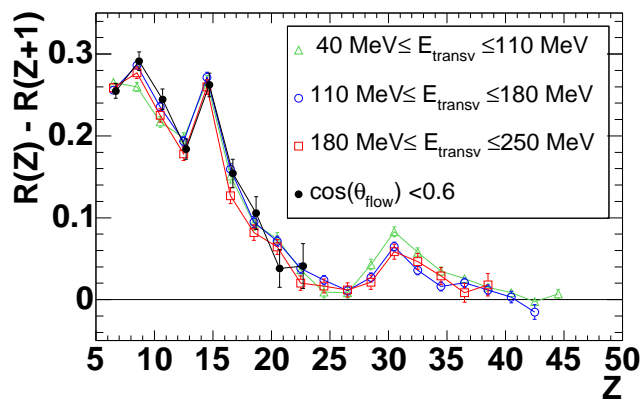}
\caption{For the reaction $^{112}Sn+^{58}Ni$ at 35 \amev, difference $R(Z)-R(Z+1)$
between the staggering ratio for even-Z values and for the following odd-(Z+1) values, for different centrality selections based on conditions on ~$\cos(\theta_\mathrm{flow}$) or E$_\mathrm{transv}$.
Adapted from Ref. \cite{cas12}.}
\label{fig22}
\end{figure}

With the CHIMERA detector two different studies have been done recently in order to elucidate the process. In the first one Casini et al. \cite{cas12} have analyzed the reaction $^{112}Sn$ + $^{58}Ni$ at 35 \amev\ at different impact parameters (from peripheral to central) looking at the staggering amplitude of the odd even effect of the IMFs fragments. The relevant result of this study is summarized in Fig. \ref{fig22}.

\begin{figure}[t]
\includegraphics[width=0.50\textwidth]{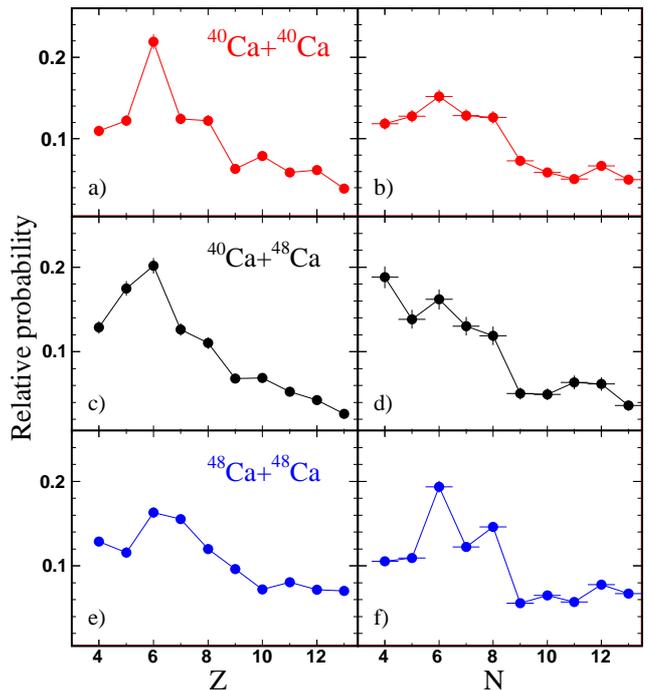}
\caption{Experimental charge $Z$ (left panels) distributions
of light fragments emitted at $10^o < \theta_{lab} < 13^o$
in the $^{40}Ca+^{48}Ca$, and $^{48}Ca+^{48}Ca$
reactions. These distributions have been normalized to
the total number of fragments emitted within the range
$4 \leq Z \leq 13$.
Experimental $N$ (right panels) distributions for the same reactions.
Vertical bars are statistical errors. Horizontal
bars are errors quoted to the identification resolution of the selected
detectors. Adapted from Ref. \cite{lom11}.}
\label{fig23}
\end{figure}

For a given bin of centrality parameter (defined using the transverse energy
and/or the $\cos(\theta_{flow})$ global variables) the staggering was studied
by an observable $R$, evaluated for a given Z, as obtained by the ratio between the experimental
yield $Y(Z)$ and an average value $\mathcal{Y}(Z)$ smoothed over the charge distribution,
following the method described in \cite{dag11}. In Fig. \ref{fig22}, the value of the incremental ratio  between the staggering ratio $R(Z)$ and $R(Z+1)$ is reported  as a function of the atomic number
$Z$ of the fragments. The different symbols refer to different selections in centrality, where $\cos(\theta_{flow})$ is used to select the most central collisions.
All the different distributions are well superimposed indicating a nearly independence
of the staggering effects from the impact parameter (from peripheral to central collisions),
supporting the argument that staggering  was mainly associated to the last steps of the decay chains.

In the second study \cite{lom11} odd-even effects in both $Z$ and $N$ distributions where analyzed for the three reactions $^{40}Ca+^{40}Ca$, $^{40}Ca+^{48}Ca$, and $^{48}Ca+^{48}Ca$ at 25 \amev, where also N (neutron number) distributions of light fragments were taken into account. Fig. \ref{fig23} shows the emission yield of light fragments detected at $10^o < \theta_{lab} < 13^o$
in complete reconstructed events.
For Z distribution (left panel) the amplitude of the staggering
is progressively smoothed with the increase of the neutron asymmetry,
from the $^{40}Ca$ + $^{40}Ca$ to the most neutron rich one $^{48}Ca$ + $^{48}Ca$.
A specular behavior is shown (right panels) for the corresponding $N$ distribution.
The effects have been qualitatively explained by assuming that in neutron-poor systems the final
charge distribution of emitted light fragments was mainly determined by odd-even effects  in the one proton separation energy of  light nuclei near the stability valley. Similar explanation was assumed for neutron rich systems where the de-excitation cascade would involve mainly neutron emissions; in such a way, the final neutron distribution of light fragments would be related to odd-even effects in the one-neutron separation energy \cite{lom11}. According to this assumption, staggering should be mainly linked with the role of pairing forces acting in the last step of the de-excitation cascade.

\begin{figure*}[ht]
\begin{center}
\includegraphics[width=0.60\textwidth]{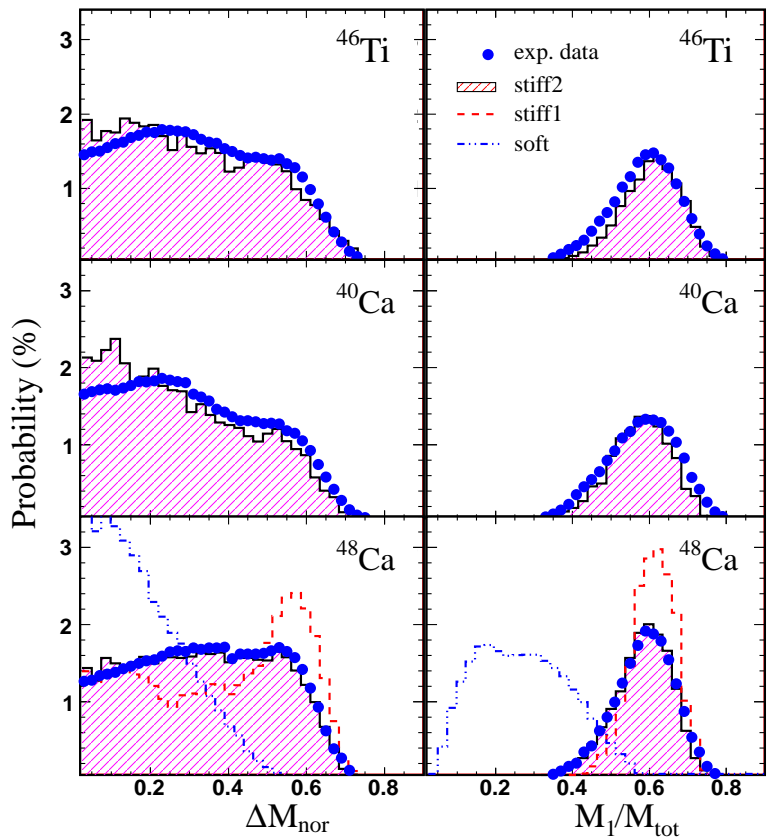}
\end{center}
\caption{Full circles: $\Delta M_{nor}$ probabilities (left panels) for reactions induced by $^{40}Ca$ on three different targets. Right panels: normalized mass of the largest fragment for the three systems.
Histograms are CoMD-II calculations with different options for the potential symmetry energy: \emph{Stiff1} ($\gamma$=1.5, dashed histograms), \emph{Stiff2} ($\gamma$=1, filled histograms) and \emph{Soft} ($\gamma=0.5$, dot-dashed histograms). \emph{Stiff1} and \emph{Soft} calculations are shown for $^{48}Ca$ target only. Adapted from Ref. \cite{amo09}.}
\label{fig24}
\end{figure*}

Understanding better the role of structure effects and dynamics in staggering phenomena is very important and it requires the development of radioactive beams in a broad range of masses and incident energy.
From an experimental point of view, detecting fragments with an accurate determination of their proton and neutron numbers is a real challenge for future experiments.
The field is fascinating involving the study of phenomena which are only poor
understood in the frame of the present status of heavy ion physics.

\section{Symmetry energy constraint from quasi-fusion reactions}
\label{sec:7}
At heavy-ions energies close to the coulomb barrier ($<$ 10 \amev) one of the most interesting
topic is the study of the dependence of decaying modes
of medium masses ($A \approx 100-130$) compound nuclei formed by fusion process \cite{lau06} from
the isospin physics. The neutron to proton ratio, $N/Z$, influences many properties of the particle decay chain of the de-excitation process,
as the level density \cite{mor12}, the staggering in the residue cross-sections, the fission barriers that are sensitive to the symmetry energy component \cite{sie85} of the semi-empirical nuclear mass formula
\cite{dan09}. Recently these aspects have been studied by the INDRA collaboration in the reactions $^{78,82}Kr$ + $^{40}Ca$ at 5.5 \amev\ at GANIL \cite{ade11} and more recently by the CHIMERA-ISODEC collaboration at LNS in the study of the reactions $^{78,86}Kr$ + $^{40,48}Ca$ at 10 \amev\ \cite{pir13}.

By increasing the incident energies from coulomb energies up to $\sim$30 \amev, in semi-central
and central collisions, the competition between binary dissipative events and heavy residue formation is expected to be sensitive to the third components of the isospin of the colliding nuclei \cite{bar05}.
An interesting early paper \cite{col98}, using a stochastic transport theory, extracted
information  about symmetry energy looking at competition between binary events and fusion in a very neutron rich induced reactions ($^{46}Ar$ +$^{64}Ni$) compared to an isospin symmetric ones ($N \approx Z$), for semi-central collisions. In this calculations it was suggested that fusion is favored for the neutron-rich systems and an asy-soft parametrization of the symmetry energy. In contrast, for the symmetric system, the more repulsive symmetry potential for protons,
shows the tendency to  decrease the coulomb repulsion,
thus favoring the fusion reaction in the asy-stiff case. Thus the interplay between the Coulomb and iso-vector interaction seems to be one of the physics key in order to understand the relative rate
of the different reaction mechanisms.

More recently, the isospin dependence of the heavy residue production cross section in incomplete
fusion reaction has been studied experimentally \cite{amo09}.
In a first experiment a $^{40}Ca$ beam (25 \amev\ incident energy) was used
to bombard  $^{48,40}Ca$ and $^{46}Ti$ targets, with CHIMERA multidetector.
Complete semi-central events were selected using criteria based on charged particle multiplicities. Particles in the event were sorted ranking the three biggest residues in an event by event analysis.
The second and third biggest fragments were gated in their velocities according to the
velocity condition ($v>v_{c.m.}$, see Refs. \cite{amo09,car12} for further details).
This last conditions was used in order to isolate massive transfer phenomena, where a very
excited source and a projectile remnant were simultaneously seen in the same event.

In Fig. \ref{fig24} the relative mass difference $\Delta M_{nor} = (M_1 - M_2)/M_{tot}$ was used
($M_1$ and $M_2$ are respectively the mas\-ses of the first and second biggest fragments and $M_{tot}$
is the total mass of the entrance system). In binary dissipative reactions we expect that
$\Delta M_{nor}$ approaches to zero, as $M_1$ and $M_2$ approach to similar values;
on the other hand, if a heavy residue is formed in the collision, the value of $\Delta M_{nor}$
should assume the largest possible value.
An inspection of the experimental results for the three different targets (Fig. \ref{fig24},
left-hand panel) reveals that the largest $\Delta M_{nor}$ is obtained for the $^{48}Ca$ neutron rich target; conversely for the two other reactions the binary deep inelastic mechanism is prevailing ($\Delta M_{nor}<0.5$). In Fig. \ref{fig24} (right-hand panel) is also shown the mass distribution of the heavy residues normalized to the total mass. The largest probability of heavy residues is observed
for the neutron rich target $^{48}$Ca.

The large neutron content in the target nucleus $^{48}$Ca in the fusion reactions induced by
$^{40}$Ca reaction favors the formation of heavy residues, because isospin asymmetry of the compound nucleus lies near the stability valley. On the other hand, the compound nucleus formed with the neutron
poor isospin symmetric systems (projectile and target) lies on the boundaries of the stability
valley (proton drip lines) \cite{amo09}.

The experimental data of Fig. \ref{fig24} were compared with CoMD-II model calculations \cite{pap01}.
In the model the symmetry interaction is derived starting from the different nucleon-nucleon (p-p, p-n, n-n) effective interactions, thus taking into account many-body correlations that involve the isospin degree of freedom \cite{pap09}. The symmetry energy is implemented with three different form factor of the symmetry potential $(\rho/\rho_0)^\gamma$ called \emph{Stiff1} (super-stiff), \emph{Stiff2} and \emph{Soft}, corresponding to $\gamma$=1.5, 1.0 and 0.5 respectively.
Comparisons with data are shown as histograms in Fig. \ref{fig24}. A good agreement with
the data was obtained using the \emph{Stiff2} option for all three systems corresponding to $\gamma=1\pm0.15$ i.e. with a linear behavior of the symmetry potential around the saturation density. The two other options adopted for the $^{48}Ca$ target produces respectively too much residues (\emph{Stiff1}, red dashed histograms) or just binary collisions (\emph{Soft}, dot-dashed histograms) where repulsive interactions are clearly prevailing.

Notice that this behavior is apparently quite different from the results of the
mean-field approach (SMF) described above in this section \cite{col98}, that
anyway is related to different reactions.
The necessity to compare predictions on symmetry energy observables as obtained by different models
of the transport dynamics is a key issues of the nowadays comparison
between theory and experimental data \cite{koh12,col10,cou11}.

\begin{figure}[ht]
\includegraphics[width=0.50\textwidth]{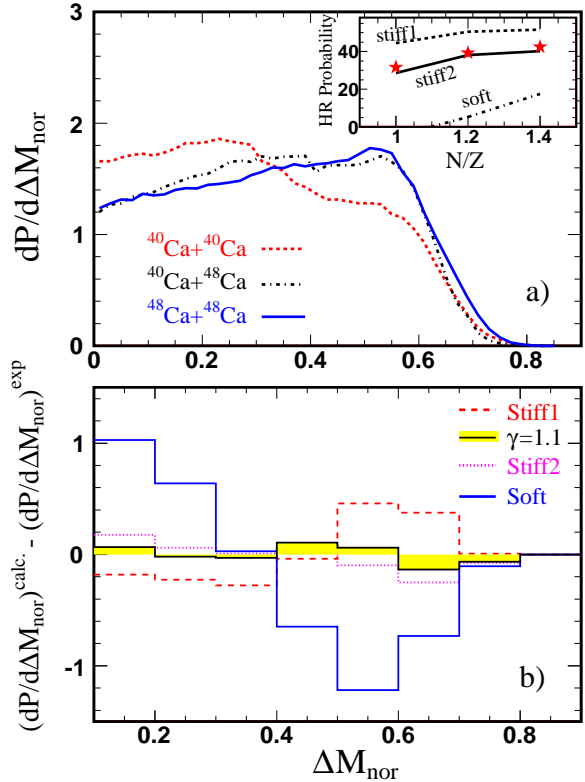}
\caption{a) $\Delta M_{nor}$ distribution for experimental data on $^{48}Ca$ + $^{48}Ca$ (blue solid line), $^{40}Ca$ + $^{40}Ca$ (red dotted line) and $^{40}Ca$ + $^{48}Ca$ (black dot-dashed line). The insert shows the probability to produce heavy residues (HR) with the condition $\Delta M_{nor}\ge 0.4$.
Stars are the experimental data. Lines are CoMD calculations. b) Deviation between calculated and experimental $\Delta M_{nor}$ distributions for different (as indicated) choices of the symmetry potential. Adapted from Ref. \cite{car12}.}
\label{fig25}
\end{figure}

In order to extend the previous study with calcium isotopes the $^{48}Ca$ + $^{48}Ca$ ($N/Z=1.4$) reaction was studied together with the mixed systems $^{40}Ca$ + $^{48}Ca$ and $^{48}Ca$ + $^{40}Ca$ at 25 \amev\ bombarding energy.
Fig. \ref{fig25}a) presents the new experimental results \cite{car12} using the
same $\Delta M_{nor}$ variable as in the previous data.

A clear evidence to increase the cross section of  heavy residue formation by increasing the neutron content of the entrance channel is clearly seen in the \ref{fig25}a). By gating  the mass of  heavy residues with the condition $\Delta M_{nor}\ge0.4$ (shown in the inset of Fig. \ref{fig25}a),
red star symbol) the maximum of the cross section was observed for the $^{48}Ca$ + $^{48}Ca$ reaction.
Calculations with the CoMD-II model confirms that data can be reproduced reasonably well using  a linear dependence \emph{Stiff2} of the symmetry potential on the
density. However, in order to  estimate better the value of the $\gamma$ parameter a linear
interpolation was adopted  over the theoretical $\Delta M_{nor}$ distributions.
The best result of $\gamma=1.1\pm0.1$ was obtained by minimizing deviations respect to the
experimental data. Fig. \ref{fig25}b) shows, for each bin of $\Delta M_{nor}$ probability distribution,
the difference between the calculated and experimental distributions for different choices of
the symmetry potential. The distribution corresponding to the minimum deviation is indicated
by the filled (yellow in colors) area histogram.

The results are coherent with the estimations of symmetry energy as obtained from the neck fragmentation mechanism. The symmetry energy was probed, in the theoretical predictions, in different region of nuclear densities: at relative low density (around $1/3\rho_0$) accordingly to SMF model for the neck fragmentation mechanism and near saturation densities (around $\rho_0\pm0.15\rho_0$)
accordingly to CoMD model predictions for fusion phenomena.

\section{Conclusions and perspectives}
\label{conclusion}
In this work we have summarized mostly of the isospin physics aspects in heavy ion reactions,
investigated with the 4$\pi$ multidetector CHIMERA in almost ten years of activities \cite{pag12};
these activities have intersected many crucial topics in heavy ion physics at medium energy.
In particular,
constraining the  nuclear symmetry energy of EOS as a function of the baryonic density
has attracted much interest in reaction studies of transport properties of the nuclear matter,
nuclear structure and astrophysics in the last decade. Important reaction mechanisms have
been reported,  using stable beams with relative large isospin asymmetries, such as neck
fragmentation and isospin diffusion, dynamical and statistical fission, heavy residue
production in semi-central collisions, formation and decay of compound nuclei,
multifragmentation of highly excited systems. The capability to detect fragments ranging
from slow-moving target-like fragments to projectile-like residues has triggered
new  heavy  ion studies. In the first part of the paper, the link among the emission
timescale of intermediate mass fragments, their isotopic composition and phase space
alignment properties and suitable transport model observables that are very sensitive
to symmetry energy has been  described. The role of isospin degree of freedom
in neutron rich and neutron poor systems has been emphasized  by a variety of interesting
results: i) neck fragmentation phenomena  where the $N/Z$ of detected IMFs
show clear signatures of the isospin dynamics; ii) the influence of isospin diffusion between
projectile and target charge asymmetric systems on isospin equilibration; iii) alignment properties and
isospin dependence of asymmetric dynamical fission of heavy projectile-like or
target-like fragments. In the second part of the paper we have shown the influence of the $N/Z$
of the colliding systems on nuclear reactions leading to single excited sources and heavy residue formation or, alternatively, to deep-inelastic binary reactions. In such studies,
a linear ($\gamma \approx 1$) behavior of the symmetry energy at sub-saturation density was deduced.
However, it was also noticed that from the analysis of isospin diffusion in Sn+Sn reactions at 35 \amev\
a moderately soft $\gamma<1$ behavior was deduced. All the obtained results, that were
consistent within the experimental uncertainties, contribute to enrich
the existing experimental constraints of the symmetry energy,
as obtained mainly by exploring isospin diffusion and neutron to proton ratio
observables \cite{tsa12}.
The researches described in this work are demanding for further investigations
in heavy ion physics with both stable and exotic beams. An important perspective has been recently envisaged  by the  possibility to couple CHIMERA detector with advanced detectors (available or under construction) covering a large portion of the available phase space in order to study particle-particle correlations \cite{ver06}. Correlations have been already investigated in CHIMERA in a first generation experiment looking for boson condensate in $^{12}C$ through the study of three alpha decay \cite{rad12}.
The Farcos (Femtoscope ARray for COrrelations and Spectroscopy) project \cite{ver13} was
envisaged in order to achieve  a compact high resolution array in both energy and angular
accuracy that can be efficiently coupled with the 4$\pi$ detector CHIMERA.
Using Farcos array in coincidence with CHIMERA is an important progress
in the topics covered in this work, and it opens new perspectives
in isospin physics with both stable and radioactive beams. The CHIMERA collaboration is also
considering  future implementations of the detector capabilities (including Farcos) by extending the measurement to neutron signals  by exploiting the characteristics of new generation plastic scintillators that are under study \cite{nim13}. In fact, exploiting both neutrons and charged particles in
a full angular coverage is a challenge for experimentalists, but, it could represent an unique experimental opportunity and important progress in heavy ion studies,
in view of the new generation exotic beams.

\begin{acknowledgement}
All the experimental results presented here were accomplished by the CHIMERA
international collaboration during several years. The authors are in particular indebted for discussions,
suggestions and support for this review work with G. Cardella, M. Colonna, I. Lombardo, M. Papa, S. Pirrone, G. Politi, F. Rizzo, P. Russotto and G. Verde.
\end{acknowledgement}
%
%
%

\end{document}